%% file: main.tex
\documentclass[10pt]{iopart}

\usepackage{graphicx}
\usepackage{iopams}
\usepackage{url}
\usepackage{color}

\begin{document}

% First page, abstract and introduction
\include{introduction_LG}

% Section 1
\include{section1_LG}

% Section 2
\include{section2_LG}

% Section 3
\include{section3_LG}

% Section 4
\include{section4_LG}

% Conclusion, aknowledgment and bibliography
%\include{conclusion_LG}
\include{conclusion_LG}

\end{document}

%% file: introduction_LG.tex
\title[Synchrotron emission from nanowire-array targets]{Synchrotron emission from nanowire-array targets irradiated by ultraintense laser pulses}

\author{B Martinez$^{1,2}$, E. d'Humi\`eres$^2$ and L. Gremillet$^1$}
\address{$^1$ CEA, DAM, DIF, F-91297 Arpajon, France}
\address{$^2$ CELIA, UMR 5107,Universit\'e de Bordeaux-CNRS-CEA, 33405 Talence, France}
\ead{bertrand.martinez@cea.fr, laurent.gremillet@cea.fr}
\vspace{10pt}

\begin{abstract}
We present a numerical study, based on two-dimensional particle-in-cell simulations, of the synchrotron emission induced during the interaction of femtosecond
laser pulses of intensities $I=10^{21}-10^{23}\,\mathrm{Wcm}^{-2}$ with nanowire arrays. Through an extensive parametric scan on the target parameters,
we identify and characterize several dominant radiation mechanisms, mainly depending on the transparency or opacity of the plasma produced by the wire expansion.
At $I=10^{22}\,\mathrm{Wcm}^{-2}$, the emission of high-energy ($>10\,\mathrm{keV}$) photons attains a maximum conversion efficiency of $\sim 10\%$
for $36-50\,\mathrm{nm}$ wire widths and $1\,\mu\mathrm{m}$ interspacing. This maximum radiation yield is found to be similar to that achieved in uniform plasma
of same average (sub-solid) density, but nanowire arrays provide efficient radiation sources over a broader parameter range. Moreover, we examine the
variations of the photon spectra with the laser intensity and the wire material, and we demonstrate that the radiation efficiency can be further enhanced by
adding a plasma mirror at the backside of the nanowire array. Finally, we briefly consider the influence of a finite laser local spot and oblique incidence angle.
\end{abstract}


\noindent{\it keywords\/}: relativistic laser-plasma interactions, particle-in-cell method, synchrotron radiation, nanowires

\submitto{\PPCF}

%\maketitle % pour ne laisser que l abstract sur le premiere page

%\ioptwocol % pour les deux colonnes

%-----------------------------------------------------------------------
 % Introduction
%----------------------------------------------------------

\section*{Introduction}

Forthcoming multi-petawatt (PW) laser systems will enable scientists to access a new regime of laser-plasma interactions where radiative and quantum
electrodynamics (QED) effects are strongly coupled with collective plasma processes \cite{RMPDiPiazza2012}. On-target laser intensities in the
$10^{22}-10^{24}\,\mathrm{Wcm}^{-2}$ range are expected to be reached at several facilities$-$CILEX-Apollon \cite{APOLLON}, PULSER \cite{APJeong2014},
ELI \cite{ELI}, Vulcan-10 PW \cite{VULCAN} and XCELS \cite{XCELS}, to name a few$-$, opening up exciting applications in fundamental and applied
research, such as radiation pressure ion acceleration \cite{PRLNaumova2009, PRLBulanov2010, PRETamburini2012}, the study of quantum radiation
reaction on laser-driven electrons \cite{PRLBlackburn2014, PRLJi2014, PoPWang2015, PRXCole2018, ArXivPoder2017}, the massive production of electron-positron
pairs through the Breit-Wheeler process \cite{PRLBell2008, PRLNerush2011, PRLRidgers2012, PoPJi2014, NCZhu2016, PREGrismayer2017, SRJirka2017},
or relativistic laboratory astrophysics \cite{HEDPLiang2013, PRLChen2015, PRLLobet2015}. One fundamental mechanism common to all these applications
is the copious generation of hard x-ray or $\gamma$-ray photons through synchrotron emission--equivalent to nonlinear inverse Compton scattering in the
strong-field regime \cite{PPCFKirk2009}. In recent years, experimental progress in this direction has been achieved by making ultrarelativistic electrons
issued from a laser wakefield accelerator collide with an intense laser pulse \cite{NPPhuoc2012, PRLChen2013, NPYan2013, NPPowers2014, PRLSarri2014, SRYu2016}.
Furthermore, the capability of this configuration in yielding efficient pair creation at laser intensities $\gtrsim 10^{23}\,\mathrm{Wcm}^{-2}$ has been
numerically \cite{PRABLobet2017} and theoretically \cite{PRABlackburn2017} examined.

The particle-in-cell (PIC) simulation technique \cite{BOOKBirdsall2004} is the most widely used tool for modeling the kinetic and collective phenomena at
play in intense laser-plasma interaction. Recently, in order to prepare for multi-PW laser experiments, much effort has been expended in enriching PIC codes
with numerical models describing synchrotron emission and multiphoton Breit-Wheeler pair production
\cite{PRLZhidkov2002, NJPTamburini2010, PPCFDuclous2011, PRECapdessus2012, PRSTABChen2013, PRLVranic2014, PREGonoskov2015,
JPCSLobet2016, PoPWallin2015, ArXivNiel2017}. Such upgraded codes are being extensively exploited to gain understanding of the radiation-modified
laser-plasma interaction in various parameter ranges. In uniform plasmas, several radiation regimes have been identified depending on the laser intensity
and plasma density. Above the relativistic critical density, an electromagnetic standing wave is formed at the laser-irradiated target front; the resulting
synchrotron radiation (referred to as skin depth emission, SDE \cite{PRLRidgers2012}) is mainly emitted in a forward-directed cone, yet remains relatively weak
(with a $\lesssim 1\%$ conversion efficiency at laser intensities $I \sim 10^{22} \,\mathrm{Wcm}^{-2}$). In relativistically near-critical or undercritical plasmas,
the radiation is predominantly emitted in the transverse (transversally oscillating electron emission, TOEE \cite{PoPChang2017}) or in the backward direction
(reinjected electron synchrotron emission, RESE \cite{PRLBrady2012}). The radiation yield has been found to be maximized in the RESE regime, with a
$\sim 1\%$ conversion efficiency predicted at $I\sim 10^{22}\,\mathrm{Wcm}^{-2}$ \cite{PoPBrady2014}. Strategies to enhance the synchrotron emission
or improve its properties have been proposed, taking advantage of preplasmas \cite{PoPBrady2014}, plasma channels \cite{PRLStark2016, APLHuang2017},
or structured targets such as gratings \cite{APLPan2015}, cone targets \cite{NCZhu2016, NJPZhu2015, OELiu2016}, clusters \cite{PoPIwata2016},
micro-plasma waveguides \cite{PRLYi2016}, or nanowire arrays \cite{QEAndreev2016, PoPLecz2017, ArXivWang2017}. The purpose of the present paper
is to further explore the potential of the latter target type for high-energy synchrotron radiation.

The realization of intense laser-driven synchrotron sources is but the latest application of nanowire (or nanotube) arrays. Originally, their use was
aimed at strongly increasing the absorption of moderately relativistic ($I \sim 10^{17}-10^{19}\,\mathrm{Wcm}^{-2}$) short-pulse lasers into fast
electrons, which can then drive bright Bremsstrahlung or x-ray line emission \cite{PoPZhao2010, LPBOvchinnikov2011, PRBMondal2011, APBIvanov2017, OHollinger2017}.
Such targets also allow for long-distance collimated transport of the fast electrons as a result of self-induced electromagnetic fields 
\cite{APLJi2010, PRLChatterjee2012, PPCFTian2014}. Moreover, fast-electron relaxation causes rapid volumetric heating and homogenization of the
nanowires, thus creating extremely hot dense matter samples. Formation of plasmas of $\sim 10^{23}\,\mathrm{cm}^{-3}$ densities and $\sim 1-10\,\mathrm{keV}$
temperatures, associated with pressures of a few Gbar, has thus been inferred by x-ray emission at $\sim 10^{19}\,\mathrm{Wcm}^{-2}$ laser
intensities \cite{NPPurvis2013, SABargsten2017}. The increased number and mean energy of the fast electrons enabled by nanowire arrays coated
on thin solid foils have also proven beneficial for accelerating ions in the target normal sheath acceleration regime \cite{SRKhaghani2017, PRLBin2018}.
One should stress, however, that these experiments raise the key issue of the laser contrast, which, if too low, may prevent the laser light from
penetrating the interwire gaps \cite{SRCristoforetti2017}.

%The main interest of nanowire-arrays lies in their ability to absorb almost all the incident laser energy. Plasmas of high densities
%$\sim 10^{23\rightarrow24}\, \mathrm{cm}^{-3}$ along with $1-10 \, \mathrm{keV}$ temperatures on lengths of $3-4 \, \mathrm{\mu m}$
%have been inferred by X-ray emission dominated by the spectral line of He-like Ni or Co \cite{NPPurvis2013,SABargsten2017} and by
%the $\mathrm{K\alpha}$ line of Zn or Co \cite{PoPZhao2010,LPBOvchinnikov2011,AIPCPSamsonova2017}. All these experiments lie in a
%laser intensity range of $I=10^{17}\rightarrow 5\times 10^{19} \,\mathrm{Wcm}^{-2}$ and raise the key issue of the laser contrast.
%Additional results have shown that a laser pre-pulse can destroy the nanostructures before the main pulse arrives and reduce the
%benefit of using nanowire-array structures \cite{PPCFCristoforetti2014,SRCristoforetti2017}. However recent progress reported
%with microwire-array targets presenting large spacings successfully irradiated at an intensity of $I=10^{21}\,\mathrm{Wcm}^{-2}$
%opens the path to the interaction of state-of-the-art lasers with nanowire-arrays. The experimentally observed properties of nano
%and micro wire arrays also include the guiding of relativistic electron beams \cite{PRLChatterjee2012}, alternative schemes for
%fast ignition \cite{PPCFTian2014}, efficient bremsstrahlung sources \cite{PRBMondal2011,EPJDJiang2014}, radiography \cite{APBIvanov2017}
%and enhanced ion acceleration \cite{PoPBeg2004,SRKhaghani2017,ArXivBin2017}.

Along with the aforementioned experimental works, a number of PIC simulation studies have examined the dependencies of the laser absorption and
fast-electron generation on the nanowire-array parameters \cite{PoPCao2010a, PoPCao2010b, QEAndreev2016, PoPLecz2017, SRCristoforetti2017}.
These works suggest that the laser absorption can reach values as high as $90\,\%$ at $I\sim 10^{19}-10^{20}\,\mathrm{Wcm}^{-2}$
and interwire spacings in the $\sim 0.1-1\,\mu\mathrm{m}$ range. The possibility of triggering betatron electron acceleration in the superimposed laser
and quasistatic fields around the wires has also been demonstrated under specific conditions (\emph{e.g.}, a $10^{19}\,\mathrm{Wcm}^{-2}$ laser
pulse irradiating 60-nm-diameter wires) \cite{QEAndreev2016, PoPLecz2017}.
% It has also been pointed out that increasing the laser intensity can prematurely damage the nanostructure, and thus reduce its benefit in terms
% of electron heating \cite{PoPCao2010b, OELi2017}. 
These trends, revealed at relatively moderate laser intensities, make nanowire arrays promising setups for developing ultraintense synchrotron
sources at extreme laser intensities, $I>10^{22}\,\mathrm{Wcm}^{-2}$. Another argument in their favor is that the strong magnetostatic fields
they give rise to (through the interplay of the fast electrons and the return current flowing inside the wires \cite{PRLKaymak2016}) may, if
sustained long enough, significantly enhance the synchrotron emission compared to that induced by the sole laser field. A similar scenario of
synchrotron radiation boosted by quasistatic fields has been numerically evidenced in a plasma channel \cite{PRLStark2016}.

In the work reported here, based on 2D simulations performed with the PIC-QED code \textsc{calder} \cite{JPCSLobet2016}, we
investigate the dominant processes of synchrotron emission and their properties as a result of the interaction of an ultraintense
($10^{21} \le I \le 10^{23}\,\mathrm{Wcm}^{-2}$), ultrashort ($30\,\mathrm{fs}$) laser pulse with a nanowire array of varying geometry.
Our paper is organized as follows. In Sect.~\ref{sec:1}, a reference scenario is presented, which considers a stand-alone nanowire array
and serves to illustrate the main stages of photon emission. In Sect.~\ref{sec:2}, we perform a parametric scan where we vary the wire
interspacing, width and atomic composition as well as the laser intensity. Our broad parameter range covers the transition from
a regime where the structure of the nanowire array is maintained during the laser irradiation to a regime where it is destroyed early in
the laser pulse, hence forming an essentially uniformized plasma. In addition, we compare the performance of nanowire arrays with that of
uniform plasmas with varying density. In Sect.~\ref{sec:3}, we show that placing a solid foil at the backside of the nanowire array can
notably increase the photon source efficiency (in case of significant laser transmission through the nanowires). Section~\ref{sec:4}
addresses the changes brought by a finite laser spot size and an oblique incidence angle. Finally, we summarize our results and suggest possible
follow-up studies.

%% file: section1_LG.tex
%-----------------------------------------------------------------------
% section 1
%-----------------------------------------------------------------------
\section{\label{sec:1} Main synchrotron emission processes in laser-nanowire-array interactions}

% section 1 - subsection 1 : numerical modelling and setup
\subsection{2D PIC simulation setup}
\label{subsec:pic_setup}

\begin{figure}[t]
\centering
\includegraphics[scale=0.8]{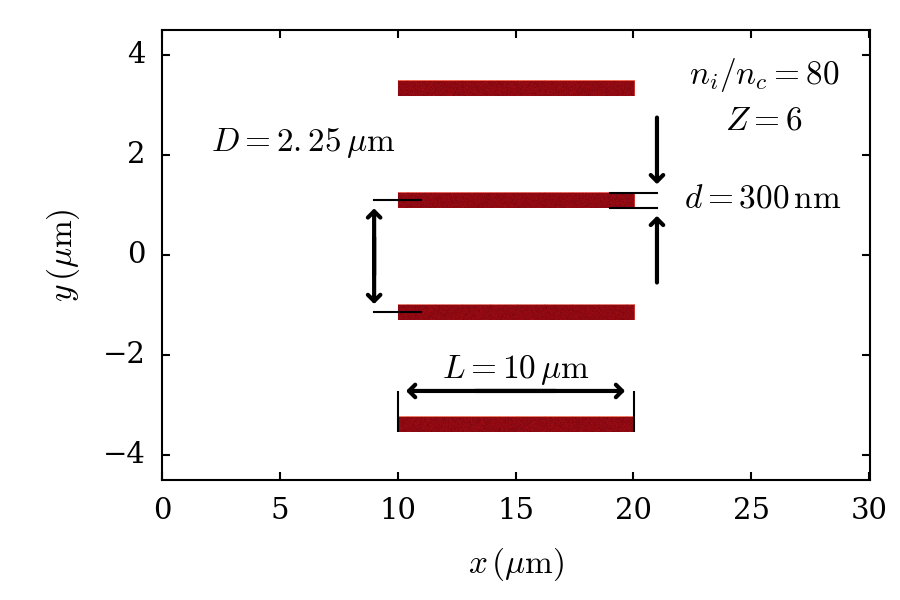}
\caption{Schematic of the reference simulation setup.}
\label{fig:figure1}
\end{figure}

In this Section, we show that the synchrotron radiation proceeds through various stages during the interaction of an ultraintense
laser pulse with a nanowire array. This is done in light of a reference 2D PIC simulation parameterized as follows. 
The laser pulse is modeled as a planar electromagnetic wave, propagating along the $x$ axis, linearly polarized along the $y$ direction
and with a central wavelength $\lambda_0=1\,\mu\mathrm{m}$. It has a Gaussian temporal profile with a FWHM duration of $30\,\mathrm{fs}$
and a peak intensity $I = 10^{22}\,\mathrm{Wcm}^{-2}$ (corresponding to a dimensionless field strength $a_0 = 85$). As depicted
in Fig.~\ref{fig:figure1}, the target consists of a periodic array of solid-density carbon nanowires. The carbon atoms, of atomic
number $Z=6$, and mass number $A=12$, are initially unionized with an atomic density $n_C = 80n_c$ ($n_c \simeq 1.1\times 10^{21}\,\mathrm{cm}^{-3}$ is
the nonrelativistic critical density). The wires have a length $L=10\,\mu\mathrm{m}$, a width (diameter in 3D) $d=0.3\,\mu\mathrm{m}$ and the
interwire spacing is $D = 2.25\,\mu\mathrm{m}$. The wire width is equal to that considered in Ref.~\cite{PRLKaymak2016},
where it was shown to give rise to strong quasistatic fields at $I=5\times 10^{21}\,\mathrm{Wcm}^{-2}$ (for circular polarization).
The absence of a substrate at the backside of the wires, which could absorb and reflect the laser pulse, allows us to isolate the
effects induced by the sole wires. The simulation domain has dimensions $L_x \times L_y = 30\,\mu \mathrm{m} \times 9\,\mu\mathrm{m}$,
with a spatial resolution $\Delta x =\Delta y =\lambda_0/210$. The temporal resolution is $\Delta t =\tau_0/314$ (where
$\tau_0 = \lambda_0/c = 3.3 \,\mathrm{fs}$ is the optical cycle) and the simulation is run over $25\,000 \Delta t$. The
boundary conditions are taken to be absorbing along $x$ and periodic along $y$ for both fields and particles, and $50$ macro-particles
per cell and per species are used. The peak of the laser pulse hits the tips of the wires at time $t=0$.

This illustrative simulation, as every other performed in this study, takes into account Coulomb binary collisions between charged particle species,
field and impact ionization and synchrotron radiation. The synchrotron module implemented in \textsc{calder} \cite{JPCSLobet2016}
combines a continuous radiation reaction model \cite{PoPSokolov2009} for electrons with a low quantum parameter ($\chi_e \leq 10^{-3}$)
and a stochastic quantum description \cite{PPCFDuclous2011} for electrons with a higher quantum parameter ($\chi_e \ge 10^{-3}$). We
recall that the electron quantum parameter, which determines the radiation characteristics, is defined as
$\chi_e = \gamma [ (\mathbf{E}_\perp + \mathbf{v}\times \mathbf{B})^2 + E_\parallel^2/\gamma^2 ]^{1/2} /E_S
\simeq \gamma \vert \mathbf{E}_\perp + \mathbf{v}\times \mathbf{B} \vert /E_S$, where $\mathbf{v}$ is the electron velocity,  $\gamma$
its relativistic factor, $\mathbf{B}$ is the magnetic field, $\mathbf{E}_\parallel$ is the electric field component parallel to $\mathbf{v}$,
$\mathbf{E}_\perp$ the electric field component normal to $\mathbf{v}$, and $E_S = m_e^2c^3/\hbar e = 1.3\times 10^{18}\,\mathrm{Vm}^{-1}$ is
the Schwinger field \cite{RMPDiPiazza2012}. The chosen threshold value between the two regimes is quite arbitrary, yet ensures that the
quantum regime is accurately described.  Pair production from Breit-Wheeler and Bethe-Heitler processes is neglected. For this reason,
and in order to reduce the computational load, the radiated photons are not advanced on the simulation grid (but their energy and emission
angle are recorded).

% section 1 - subsection 2 : three contributions through an example
\subsection{Typical dynamics of the laser-nanowire-array interaction and its associated synchrotron emission}
\label{subsec:reference_simulation}

\begin{figure}
\centering
\includegraphics[scale=0.8]{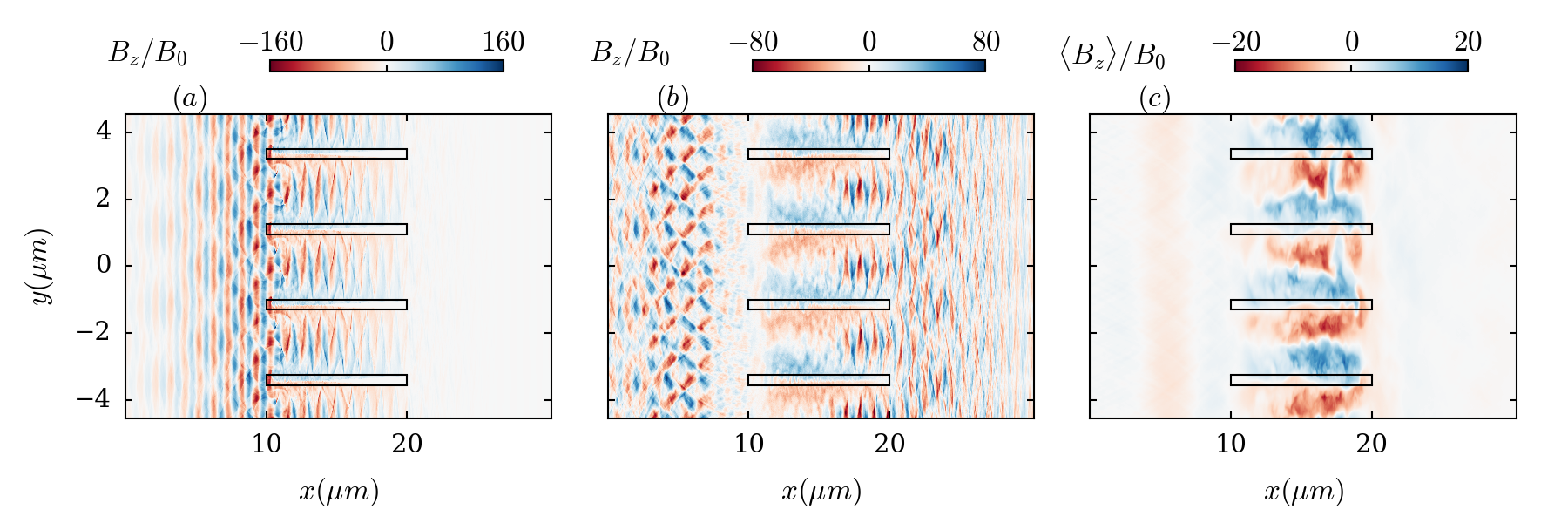}
\caption{Maps of the magnetic field $B_z$ (normalized to $B_0=m_e \omega_0/e \simeq 1.1\times10^4 \,\mathrm{T}$) at three different times: 
(a) $t=8\,\mathrm{fs}$ (during plasma filling of the interstices), (b) $t= 40 \,\mathrm{fs}$ (after the left-hand side of the plasma-filled interstices
have become opaque to the laser) and (c) $t=167\,\mathrm{fs}$ (final simulation time). The peak of the laser pulse hits the wire tips at $t=0$. Panel (c)
displays the magnetostatic field, $\langle B_z \rangle$, averaged over an optical cycle. The black rectangles plot the initial location of the wires.}
\label{fig:figure2a_2c}
\end{figure}
\begin{figure}
\centering
\includegraphics[scale=0.8]{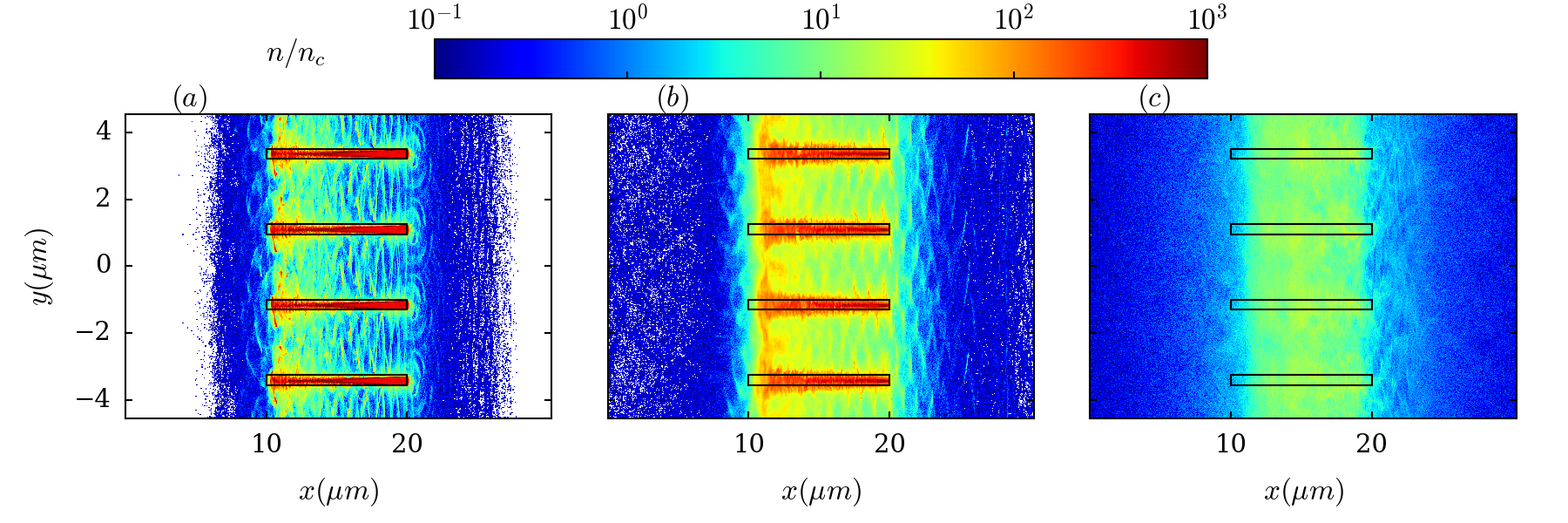}
\caption{Maps of the electron density $n_e$ (normalized to the nonrelativistic critical density $n_c \simeq 1.1\times 10^{21}\,\mathrm{cm}^{-3}$) at
(a) $t=8\,\mathrm{fs}$ and (b) $t= 40 \,\mathrm{fs}$. Panel (c) displays the ion density $n_i$ at $t=167\,\mathrm{fs}$ (final simulation time).}
\label{fig:figure3a_3c}
\end{figure}

Figures~\ref{fig:figure2a_2c}(a-c) display maps of the magnetic field ($B_z$) at three successive times, visualizing the penetration
of the laser wave through the interwire gaps and the generation of quasistatic fields. The magnetic field is normalized to
$B_0 = m_e \omega_0 /e = 1.1\times 10^4 \,\mathrm{T}$ (where $m_e$ is the electron mass, $e$ is the elementary charge, and $\omega_0$
is the laser angular frequency). The expansion dynamics of the wires is illustrated by the electron and ion density maps shown in
Figs.~\ref{fig:figure3a_3c}(a-c). At the beginning of the interaction, the electrons are pulled over a $\sim 1\,\mu\mathrm{m}$ distance from the
wire surface by the $E_y$ component of the laser field, and accelerated in the forward direction by its $B_z$ component. As a result, the interwire
gaps are filled with a population of energized electrons bunched at the laser wavelength $\lambda_0$.  Figure~\ref{fig:figure3a_3c}(a) is recorded
shortly after the on-target laser peak ($t=+8\,\mathrm{fs}$), at which time the electron density in the interstices near the tips of the wires is of
$\sim 50n_c$, \emph{i.e.}, approaches the relativistic critical density $n_{cr} \simeq a_0 n_c$ (see also Sec.~\ref{sec:comp_unif_targets}).
Figure~\ref{fig:figure2a_2c}(a) shows that, up to this time, the interstices have remained (partially) transparent to the laser wave. The hot-electron
current flowing in the interstices induces a magnetostatic field that is screened inside the wires by a return current carried by bulk electrons
(of density $n_e \simeq Zn_C =480n_c$). The amplitude of this field can be estimated by noting that the laser-accelerated electrons are initially
extracted from a layer of thickness $\delta_{acc} \simeq a_0 (n_c/n_e) c/\omega_0 \simeq 30\,\mathrm{nm}$ (assuming immobile ions and a balance
between the transverse laser and space-charge fields). These electrons generate a magnetostatic field of normalized
strength $\langle B_z \rangle /B_0 \simeq \langle v_x /c \rangle (n_e/n_c) \delta_{acc} \omega_0/c \simeq a_0 \langle v_x/c \rangle$,
with $\langle v_x \rangle \simeq c$ the mean longitudinal fast-electron velocity. One therefore expects the strength of the self-induced
magnetostatic field to be comparable with that of the laser field, in agreement with the maximum value $\langle B_z \rangle \simeq 0.7 a_0$
measured at the laser peak. At laser intensities (resp. wire width) high (resp. small) enough that $\delta_{acc} \gtrsim d/4$,
the number of electrons remaining inside the wires becomes lower than those expelled by the laser, so that current balance between
the forward-moving hot electrons and the backward-moving core electrons \cite{PRLKaymak2016} can no longer be maintained in the vicinity
of a wire. In the planar-wave case under consideration, this leads to $\langle B_z \rangle$ dropping with decreasing $d \lesssim 4\delta_{acc}$,
from $\langle B_z \rangle /B_0 \simeq (n_e/n_c) d \omega_0/4c$ down to zero in the fully depleted regime ($d \lesssim 2\delta_{acc}$).

\begin{figure}
\centering
\includegraphics[scale=0.8]{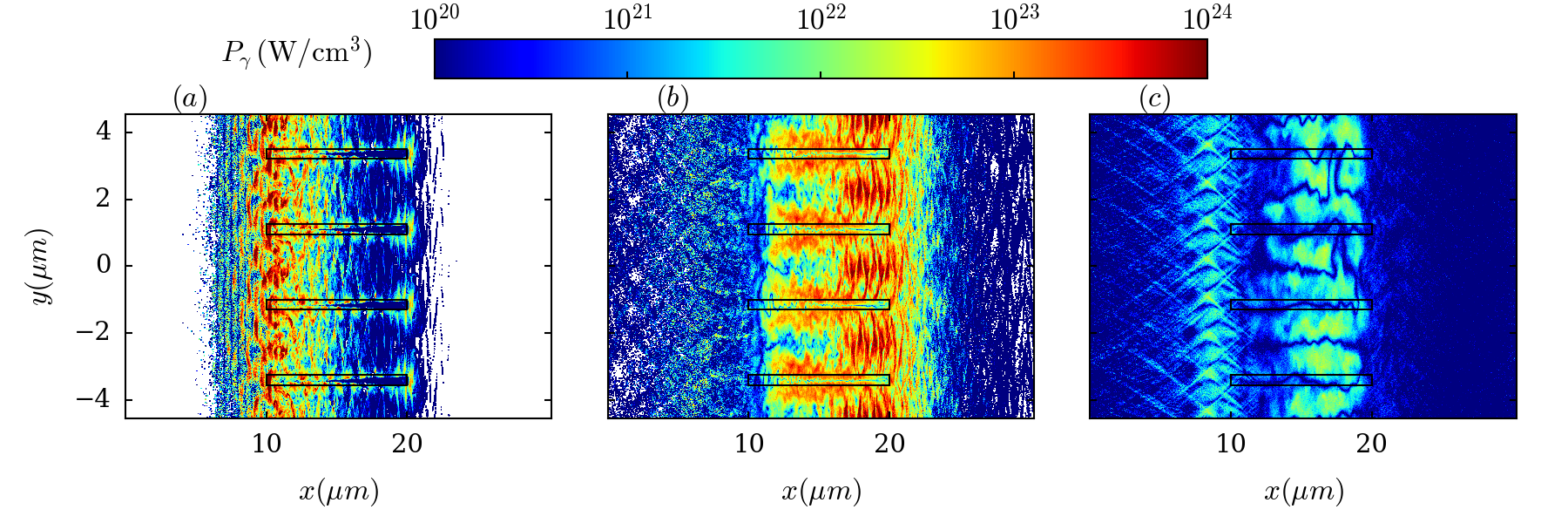}
\caption{Maps of the radiated power density, $P_\gamma$, at three different times: (a) $t=8\,\mathrm{fs}$ (during plasma filling
of the vacuum gaps), (b) $t= 40 \,\mathrm{fs}$ (after the left-hand side of the plasma-filled interstices have become opaque to the laser) and
(c) $t=167\,\mathrm{fs}$ (final simulation time).}
\label{fig:figure4a_4c}
\end{figure}
\begin{figure}[h]
\centering
\includegraphics[scale=0.8]{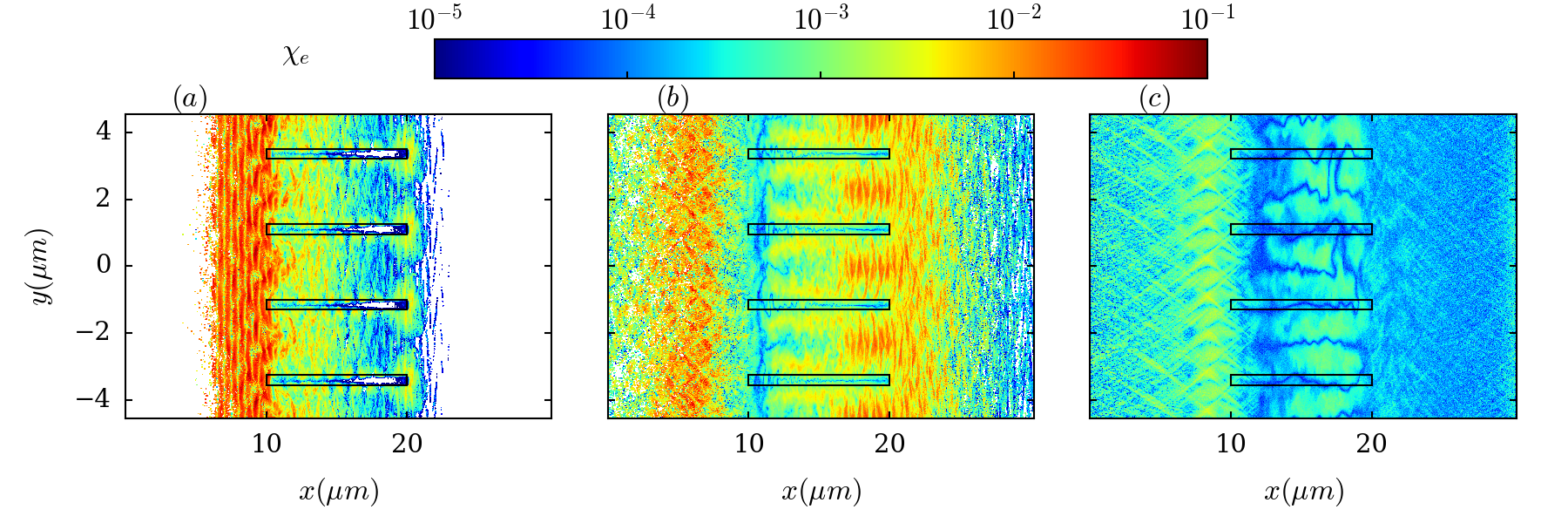}
\caption{Maps of the electron quantum parameter, $\chi_e$, at three different times: (a) $t=8\,\mathrm{fs}$, (b) $t= 40 \,\mathrm{fs}$ 
and (c) $t=167\,\mathrm{fs}$ (final simulation time).}
\label{fig:figure5a_5c}
\end{figure}

The magnetostatic field tends to deflect inwards the bulk electrons, resulting in the pinching of the wire cores \cite{PRLKaymak2016}.
This transverse magnetic compression occurs early in time, as shown in Fig.~\ref{fig:figure3a_3c}(a) where one can note a contraction of the
wires compared to their initial position (solid black lines). Simultaneously, the space-charge sheath field $\langle E_y \rangle/B_0c \simeq (n_e/n_c) \delta_{acc}\omega_0/c$
transversely accelerates the ions from the outer wire regions, and hence an increasingly dense plasma progressively fills up the interwire gaps.
At $t=+40\,\mathrm{fs}$, the bulk electrons have expanded enough to form in the wire interstices a relativistically overcritical plasma ($n_e \ge n_{cr}$)
opaque to the laser light [Fig.~\ref{fig:figure3a_3c}(b)]. This causes the splitting of the laser pulse into a transmitted part and a reflected one,
as seen in Fig.~\ref{fig:figure2a_2c}(b). The density modulations at the plasma surface arising from the incomplete homogenization of the wires account
for the reflection interference pattern seen in front of the target. Given the relatively large interwire spacing considered here, the laser
transmission across the target is significant ($\simeq 13\,\%$).  Figure~\ref{fig:figure3a_3c} shows that, by $t=+167\,\mathrm{fs}$ (about $85\,\mathrm{fs}$
after the laser pulse has exited the simulation domain), the nanostructure has been completely homogenized, the ion density then tending to the average
density $n_id/D= 11n_c$. 

Figure~\ref{fig:figure2a_2c}(c) plots the quasistatic magnetic field $\langle B_z \rangle$, averaged over an optical cycle, at $t=167\,\mathrm{fs}$.
It demonstrates the relatively slow decay of the magnetostatic modulations sustained by the homogenized target electrons. At this instant, 
these modulations have a strength $\langle B_z \rangle \simeq 15 B_0 \simeq 0.18 a_0$, which remains an appreciable fraction of the laser
field, and a typical variation length of $\sim 0.5\,\mu\mathrm{m}$, leading to magnetization of electrons with up to $\sim 25\,\mathrm{MeV}$ energies,
and therefore of the vast majority of the plasma electrons, of mean energy $\langle \gamma \rangle m_e c^2 \simeq 12\,\mathrm{MeV}$.

\begin{figure}
\centering
\includegraphics[scale=0.8]{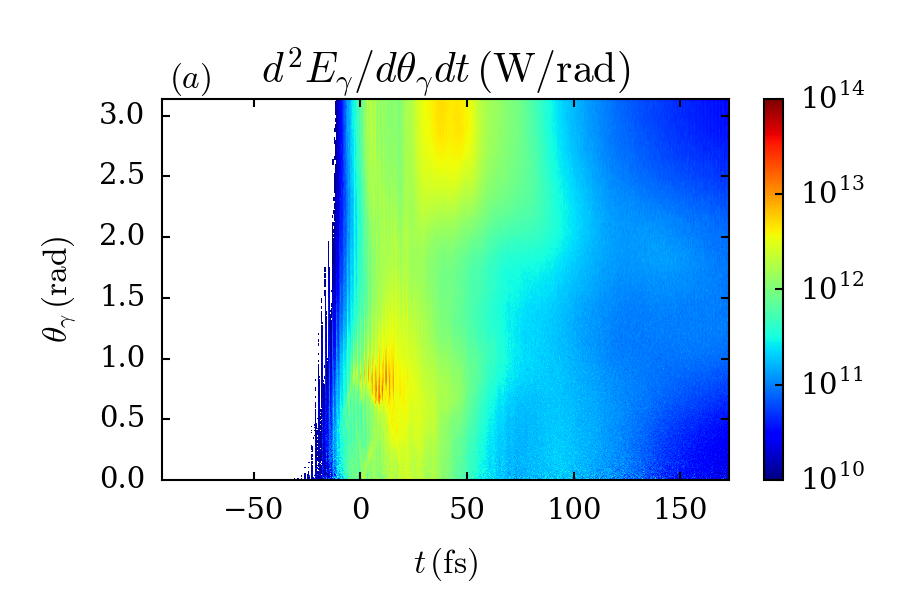}
\includegraphics[scale=0.8]{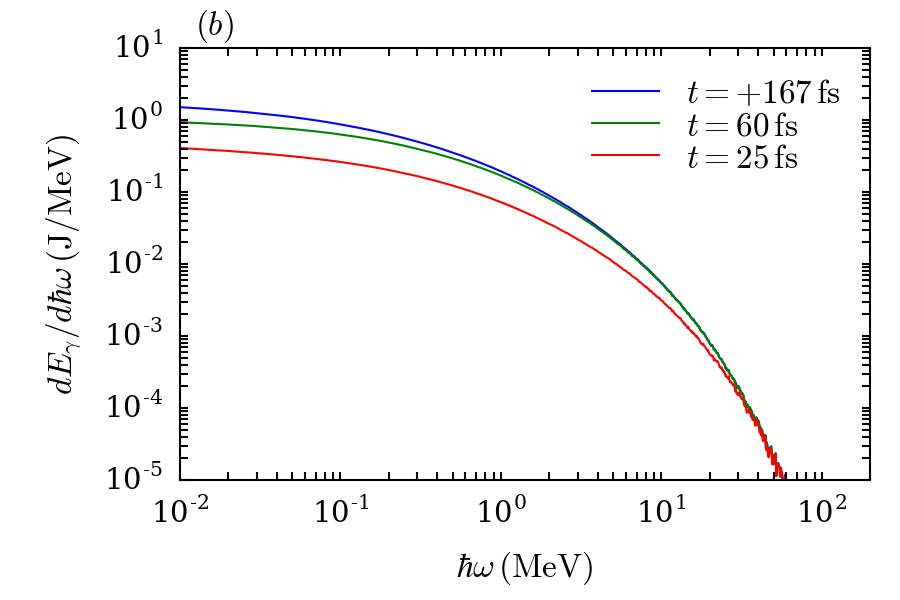}
\caption{(a) Time evolution of the angle-resolved radiated power and (b) photon energy spectra integrated from the start of the interation
up to three different times: $t=+25\,\mathrm{fs}$ (red curve), $+60\,\mathrm{fs}$ (green curve) and $+167\,\mathrm{fs}$ (blue curve).
Angles in (a) are defined as $\theta_\gamma = \arccos(k_{\gamma,x}/k_\gamma) \in (0,\pi)$.}
\label{fig:figure6a_6b}
\end{figure}

Let us now examine the synchrotron emission that takes place during and after the laser-nanowire interaction. To provide insight into the radiative processes,
we plot the spatial distribution of the averaged (over the local electron distribution) radiated power density in Figs.~\ref{fig:figure4a_4c}(a-c) and of
the averaged electron quantum parameter, $\chi_e$ in Figs.~\ref{fig:figure5a_5c}(a-c), at the same times as in Figs.~\ref{fig:figure2a_2c}(a-c). We remind
that the power radiated by a single electron can be expressed as $P = \left(2/3\right)\alpha_f m_e c^2 \chi_e^2 g(\chi_e) /\tau_C$, with $\tau_C=\hbar/m_e c^2$
the Compton time, $\alpha_f$ the fine structure constant, and $g(\chi_e)$ a quantum correction \cite{PPCFKirk2009}. The scaling $P\propto \chi_e^2$ is a good
approximation in the classical regime ($\chi_e \lesssim 0.05$). These maps will help analyze the time evolution of the angle-resolved radiated power and the
photon energy spectra (integrated over different time intervals) plotted in Figs.~\ref{fig:figure6a_6b}(a,b). The angle $\theta_\gamma$ denotes the angle of
the photon momentum $\mathbf{k}_\gamma$ relative to the laser axis ($x$), \emph{i.e.}, $\theta_\gamma=\arccos(k_{\gamma,x}/k_\gamma)\in (0,\pi)$.

Figure~\ref{fig:figure6a_6b}(a) indicates that the emission initially occurs in the laser direction ($\theta_\gamma =0$) with an increasingly broad angular
distribution. As expected, the emission strongly increases at the laser peak, and is at its brightest in the time period $5 \lesssim t \lesssim 20\,\mathrm{fs}$.
Figure~\ref{fig:figure6a_6b}(b) shows that the spectrum radiated from the start of the interaction up to $t=+25\,\mathrm{fs}$ extends to
$\hbar \omega_\mathrm{max} \simeq 60\,\mathrm{MeV}$, and makes up $\sim 40\,\%$ of the total radiated energy. The radiated power is then contained in
a forward cone of $\sim 45^\circ$ half angle and is modulated at twice the laser frequency. This oscillation is typical of the synchrotron radiation
from a relativistically overdense plasma layer in the SDE regime \cite{PPCFBrady2013}.
Consistently, Fig.~\ref{fig:figure4a_4c}(a), recorded at $t=+8\,\mathrm{fs}$, shows that the emission then mainly occurs at the front side of the plasma
(with electron density $n_e \simeq 10-50 n_c$) filling the wire gaps, where relatively high values $\chi_e \simeq 5\times 10^{-2}-10^{-1}$ are found.
Deeper into the array ($15\le x \le 20\,\mu\mathrm{m}$), the more dilute, $\lambda_0$-periodic electron bunches that move along the laser wave present
a weaker quantum parameter, $\chi_e <10^{-2}$ (due to compensating electric and magnetic forces), and hence emit little energy. Note, however, the relatively
bright synchrotron spots at the right-hand tips of the wires, where space-charge fields deflect the electrons at an angle to the laser direction, hence increasing
their quantum parameter (see below).  

Figure~\ref{fig:figure6a_6b}(a) reveals that a secondary emission burst occurs in the time interval $25 \lesssim t \lesssim 60\,\mathrm{fs}$, when
the transmitted laser pulse travels across the target backside. In contrast to the first emission burst, this emission takes place in the backward
direction ($\theta_\gamma \simeq \pi$). It originates from the interaction of the transmitted part of the laser pulse with the fast electrons reflected
at the target backside by the space-charge field. Such a counterpropagating geometry maximizes the quantum parameter $\chi_e \simeq 2\gamma E_\perp /E_S$
(where $E_\perp$ is the laser electric field). This mechanism is supported by Fig.~\ref{fig:figure4a_4c}(b), which shows a volumetric emission between
(and near the backside of) the wires, where $\chi_e$ values of $\sim 10^{-2}$ are reached [Fig.~\ref{fig:figure5a_5c}(b)]. About $45\,\%$ of the synchrotron
yield is radiated during this stage (with maximum photon energies $\sim 60\,\mathrm{MeV}$, similar to those in the primary stage). Of course, this
phenomenon will be altered in the presence of a substrate coated at the target backside (see Sec.~\ref{sec:3}). We note that relatively high $\chi_e$
values ($\sim 10^{-2}$) are also reached in the dilute plasma formed in front of the target, yet the electron density, $n_e \simeq 0.1n_c$
[Fig~\ref{fig:figure3a_3c}(b)] is there too low to yield significant emission.

Following the laser irradiation ($t \gtrsim 60\,\mathrm{fs}$), the radiated power strongly drops, yet, in similar fashion to Ref.~\cite{PRLStark2016},
the remaining magnetostatic fields can sustain additional radiation [Fig.~\ref{fig:figure6a_6b}(a)]. Figure~\ref{fig:figure5a_5c}(c) thus indicates
that, at $t=+167\,\mathrm{fs}$, $\chi_e$ attains values $\sim 10^{-3}$ in the magnetic modulations. The weaker power radiated at such low $\chi_e$ values
[Fig.~\ref{fig:figure4a_4c}(c)] is partially compensated for by the longer duration of this emission stage, which makes up $\sim 15\,\%$ of the
total yield in the time period $60\lesssim t \lesssim 167\,\mathrm{fs}$ [Fig.~\ref{fig:figure6a_6b}(b)]. Since the magnetostatic fields build up early
in the laser irradiation, their contribution is \emph{a priori} not limited to the final times of the simulation. Yet their effect is initially
mitigated by the transverse electrostatic field ($\langle E_y \rangle$) around the wires, which tends to weaken the quantum parameter; as the wires
radially expand and mix [Fig.~\ref{fig:figure3a_3c}(c)], however, $\langle E_y \rangle$ diminishes and becomes small compared to $\langle B_z \rangle$,
so that  $\chi_e \simeq \langle \gamma \rangle \langle B_z \rangle c/E_S$. At $t=167\,\mathrm{fs}$, we have $\langle B_z\rangle \simeq 15B_0$ and
a mean electron energy $\langle \gamma_e \rangle \simeq 23$ in the expanded plasma, which implies $\chi_e \simeq 8\times 10^{-4}$, consistent with
Fig.~\ref{fig:figure5a_5c}(c).

Our reference simulation has allowed us to pinpoint important processes affecting the synchrotron radiation in the interaction
of a $10^{22}\,\mathrm{Wcm}^{-2}$ femtosecond laser pulse with a nanowire array of micron-scale interspacing. We will now examine
the dependencies of the emission on the wire and laser parameters.
%The early filling of the wires enables to split the laser into a reflected and a trasmitted part. The reflection interacts with an elongated preplasma
%providing fast electrons and leads to a surface emission. The transmission is trapped inside the wires and fosters a volumetric emission.
%The formation of strong quasi-static fields driven by the compensating return current which amplitude is of the order of the laser one can boost
%and sustain the radiation on a longer time scale than the laser duration.

%% file: section2_LG.tex
%-----------------------------------------------------------------------
% section 2
%-----------------------------------------------------------------------
\section{\label{sec:2}Parametric scan on the nanowire parameters and the laser intensity}

In the following, we explore the dependency of the angle-energy spectra of the synchrotron radiation on the nanowire spacing ($D$),
width ($d$) and material ($Z$), as well as on the laser intensity ($I$). Except for the varied parameter, the numerical setup is
identical to that presented in Sec.~\ref{subsec:pic_setup}. Our parametric scan will encompass various regimes of synchrotron
radiation, which will be interpreted in light of the processes revealed in the reference case of Sec.~\ref{subsec:reference_simulation}
and previous simulation works \cite{PoPCao2010a, PRLBrady2012, QEAndreev2016,PoPLecz2017}.

% section 2 - subsection 1 : spacing dependence
\subsection{\label{subsec:variation_spacing}Variation of the interwire spacing: from forward to backward directed radiation}

\begin{figure}[t]
\centering
\includegraphics[scale=0.8]{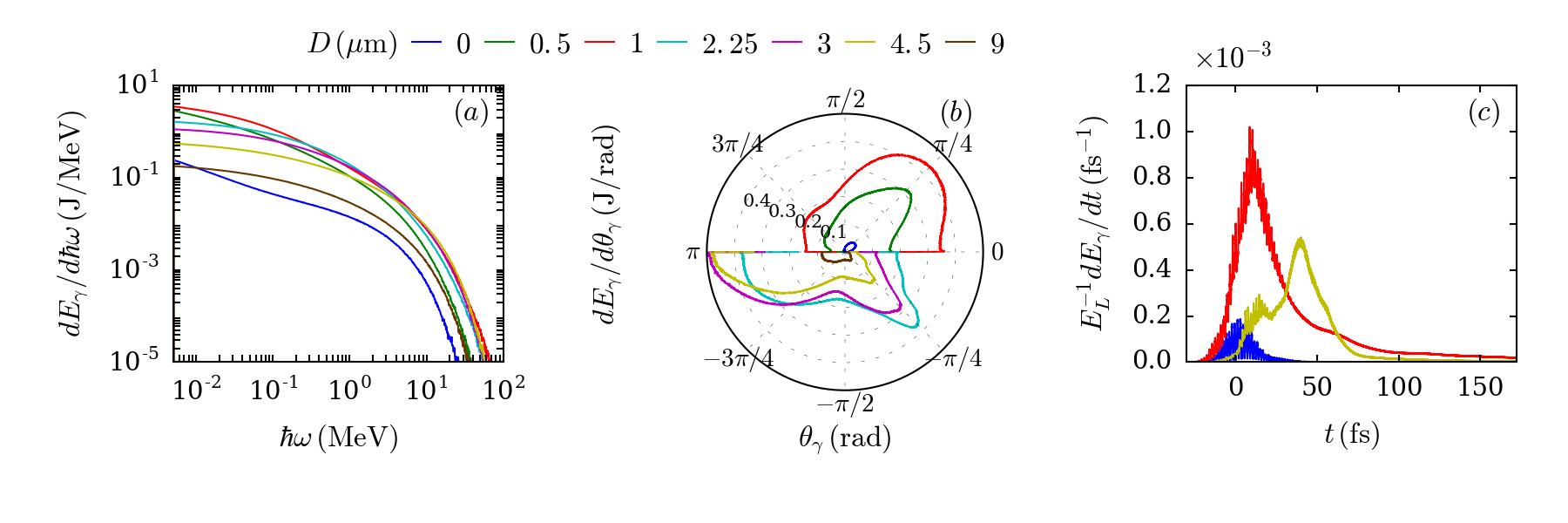}
\caption{Variations of the synchrotron emission with the interwire spacing $D$: (a) energy spectra, (b) angle-resolved radiated energy and
(c) time-resolved radiated power (and normalized to the total laser energy $E_L$). Each color represents a different value of $D$
(in $\mu\mathrm{m}$ units) as indicated in the legend of panel (a). Angles in (b) are defined as $\theta_\gamma = \arccos(k_{\gamma,x}/k_\gamma) \in (0,\pi)$
and the resulting angular distribution is symmetrized with respect to $\theta_\gamma = 0$. All plotted quantities are integrated over
the simulation domain.}
\label{fig:figure7a_7c}
\end{figure}

In our simulations, the interwire spacing has been varied over the set of values $D \in [0, 0.5, 1, 2.25, 3, 4.5, 9]\,\mu\mathrm{m}$.
Note that $D=0\,\mu\mathrm{m}$ corresponds to a planar target. The chosen values exactly divide the transverse size of the
domain ($L_y=9\,\mu\mathrm{m}$) so as to keep the periodic condition valid. The other target parameters are set to $d=0.3\,\mu\mathrm{m}$,
$L=10\,\mu\mathrm{m}$, $Z=6$ and the laser intensity is $I=10^{22}\,\mathrm{Wcm}^{-2}$.

The energy-resolved photon spectra recorded for various interwire spacings are plotted in Fig.~\ref{fig:figure7a_7c}(a). We see
that the cutoff photon energy weakly varies for $1 \leq D \leq 4.5\,\mu\mathrm{m}$, where it reaches a maximum value
$\hbar \omega_\mathrm{max} \simeq 50\,\mathrm{MeV}$, approximately twice that found at uniform density ($\simeq 23\,\mathrm{MeV}$).
Figure~\ref{fig:figure7a_7c}(b), which displays the angle-resolved enery spectra, shows a transition from a mainly forward-directed
emission at $D \le 1 \,\mu\mathrm{m}$  to an increasingly backward-directed emission at larger spacings. The two lobes of emission
found at $D \ge 2.25\,\mu\mathrm{m}$ around the directions $\theta_\gamma \simeq 45^\circ$ and $\theta \simeq 180^\circ$ originate from
the same mechanisms discussed in Sec.~\ref{subsec:reference_simulation}.
In particular, we emphasize that the backward emission follows from the electrons refluxing in the $-x$ direction and colliding head-on
with the transmitted part of the laser pulse. This results in a secondary backward-directed $\gamma$-ray burst after the primary (and weaker)
forward-directed burst. This is evidenced in Fig.~\ref{fig:figure7a_7c}(c) where is plotted the time evolution of the radiated power: 
the curve at $D=4.5\,\mu\mathrm{m}$ presents two distinct emission peaks, the second, brighter one taking place at $t\simeq 40\,\mathrm{fs}$,
\emph{i.e.}, as the laser pulse exits the target. 

\begin{figure}[t]
\centering
\includegraphics[scale=0.8]{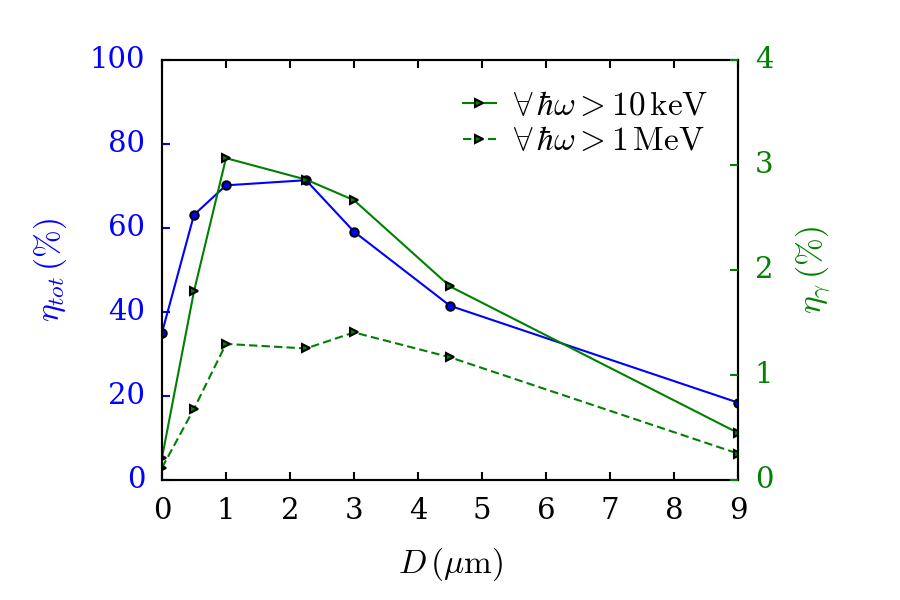}
\caption{Variations with the interwire spacing ($D$) of the total absorbed laser energy fraction ($\eta_{tot}$, blue circles) and radiation
conversion efficiency ($\eta_\gamma$, green triangles). The radiation conversion efficiency is computed for two photon energy thresholds: 
$\hbar \omega \geq 10\,\mathrm{keV}$ (green solid) and $\hbar \omega \geq 1\,\mathrm{MeV}$ (green dashed). All quantities are integrated over the 
simulation duration.}
\label{fig:figure8}
\end{figure}
   
At narrower spacings ($D \leq 1\,\mu\mathrm{m}$), the interstices fill up with opaque plasma increasingly early before the laser pulse maximum.
Looking at the increase in the instantaneous laser reflectivity, we find that the transparency-opacity transition occurs at $\tau_f \simeq -8\,\mathrm{fs}$
for $D=0.5\,\mu\mathrm{m}$ and $\tau_f \simeq 3 \,\mathrm{fs}$ for $D=1\,\mu\mathrm{m}$. The energy fraction and mean intensity of the transmitted
light then diminishes with decreasing $D$, which greatly weakens the aforementioned backward emission mechanism. At $D=2.25\,\mu\mathrm{m}$,
about $13\,\%$ of the laser energy is transmitted, and this fraction becomes negligible for $D \leq 1\,\mu\mathrm{m}$. The time history of the radiated power
at $D=1\,\mu\mathrm{m}$, plotted in Fig.~\ref{fig:figure7a_7c}(c), thus presents a single maximum, ocurring at $t \simeq 10\,\mathrm{fs}$, just after
the overdense plasma filling of the vacuum gaps. The primary radiation burst observed at $D = 4.5\,\mu\mathrm{m}$ occurs approximately at the
same time: both signals exhibit a $2\omega_0$ modulation, characteristic of SDE in an overcritical plasma \cite{PPCFBrady2013}. The photons are
then emitted in a large forward cone, as seen in the upper part of Fig.~\ref{fig:figure7a_7c}(b).

As pointed out in the Introduction, the interest for nanowire targets as potentially efficient radiation sources arose from their well-established
capability in yielding high laser absorption fractions. Since the latter usually translate in large numbers of energetic electrons, it is tempting
to predict that the laser absorption and radiation yield are correlated. To check this scenario, we plot in Fig.~\ref{fig:figure8} the variations of
the total absorbed laser energy fraction ($\eta_{tot}$, defined as the energy absorbed by all the particle and photon species, normalized to the
laser energy) and the laser-to-photon energy conversion efficiency ($\eta_\gamma$) with the interwire spacing. To discriminate between the contributions of
the `low' and `high' energy photons in the radiation yield, the green solid and dashed $\eta_\gamma$ curves are computed applying lower-energy cutoffs
$\hbar \omega = 10\,\mathrm{keV}$ and $1\,\mathrm{MeV}$, respectively. We note that the laser absorption rises from $\sim 35\,\%$ at uniform solid density
to $\sim 70\,\%$ at $D=2.25\,\mu\mathrm{m}$, with a plateau above $\sim 60\,\%$ in the range $1 \le D \le 3\,\mu\mathrm{m}$. While the $\eta_{tot}$ and
$\eta_\gamma$ curves look similar, a few quantitative differences are discernible. Both $\eta_\gamma$ curves starting from very low values ($\sim 0.2\,\%$
for $\hbar \omega \ge 10\,\mathrm{keV}$ and $\sim 0.1\,\%$ for $\hbar \omega \ge 1\,\mathrm{MeV}$) at uniform solid density, they present a steeper
rise at low $D$ values ($\le 1\,\mu\mathrm{m}$) than $\eta_{tot}$. Also, $\eta_\gamma$ attains its maximum ($\sim 3\,\%$, for $\hbar \omega \ge 10\,\mathrm{keV}$) 
at $D=1\,\mu\mathrm{m}$, lower than the value optimizing $\eta_\gamma$.  For $\hbar \omega \ge 1\,\mathrm{MeV}$, we find $\eta_\gamma \simeq 1\,\%$
in a broader range of interwire spacings, $1\leq D \leq 4.5\,\mu\mathrm{m}$, with a weakly pronounced optimum at $D=3\,\mu\mathrm{m}$.

The overall evolution of the total laser absorption, as depicted in Fig.~\ref{fig:figure8}, is consistent with the results obtained in Ref.~\cite{PoPCao2010a}
at lower laser intensity ($I=5\times 10^{19} \,\mathrm{W/cm}^{2}$) and in the sub-micron range $0.24 \le D \le 0.8\,\mu\mathrm{m}$ (with $d=0.16\,\mu\mathrm{m}$). 
In our work, by considering larger interwire spacings, we allow greater fractions of the laser light to be transmitted through the target, thus enabling
the secondary radiation burst at the target backside discussed above. Moreover, in the laser intensity range $10^{18}-10^{21}\,\mathrm{Wcm}^{-2}$, 
it is commonly believed that increasing the wire spacing enables the electrons to reach higher energies~\cite{PRLJiang2016, PoPLecz2017, SRCristoforetti2017}.
Our results partially corroborate this behavior at $I=10^{22}\,\mathrm{Wcm}^{-2}$: the mean energy of the electrons above $511\,\mathrm{keV}$ is
found to increase from $\langle E_e \rangle = m_ec^2 \langle \gamma \rangle \simeq 5 \,\mathrm{MeV}$ at $D=0$ to $\langle E_e \rangle \simeq 15 \,\mathrm{MeV}$
at $D=2.25\,\mu\mathrm{m}$. At larger spacings, $2.25\leq D \leq 9\,\mu\mathrm{m}$, the mean hot-electron energy is found to saturate at
$\langle E_e \rangle \simeq 20\,\mathrm{MeV}$, relatively close to the ponderomotive scaling
$\langle E_e \rangle \simeq m_ec^2 \left(\sqrt{1+a_0^2/2}-1\right) \simeq 30\,\mathrm{MeV}$ \cite{PRLWilks1992, PRLFiuza2012}.

To summarize, we have identified two distinct regimes of synchrotron radiation by varying the interwire spacing. For narrowly spaced wires
($D\le 1\,\mu\mathrm{m}$), the vacuum gaps rapidly fill up with overdense plasma before the on-target laser peak, causing the emission to be
concentrated at the target front and mainly forward directed, similarly to what occurs in a uniform overdense plasma. At larger interwire
spacings ($D\ge 2.25\,\mu\mathrm{m}$), this mechanism is progressively superseded by an additional emission taking place at the target backside,
which results from the interaction of the transmitted laser light with the refluxing fast electrons. This backward-directed emission is distinct
from the RESE mechanism highlighted in Ref.~\cite{PRLBrady2012}, which occurs at the moving laser front in relativistically underdense plasmas.
To achieve the dilute plasma conditions required by the latter mechanism during the laser pulse, the wire width must be reduced, as is done in
the next Section.

% section 2 - subsection 2 : diameter dependence
\subsection{\label{sec:variation_width}Variation of the wire width: from RESE to SDE, through TOEE}

\begin{figure}
\centering
\includegraphics[scale=0.8]{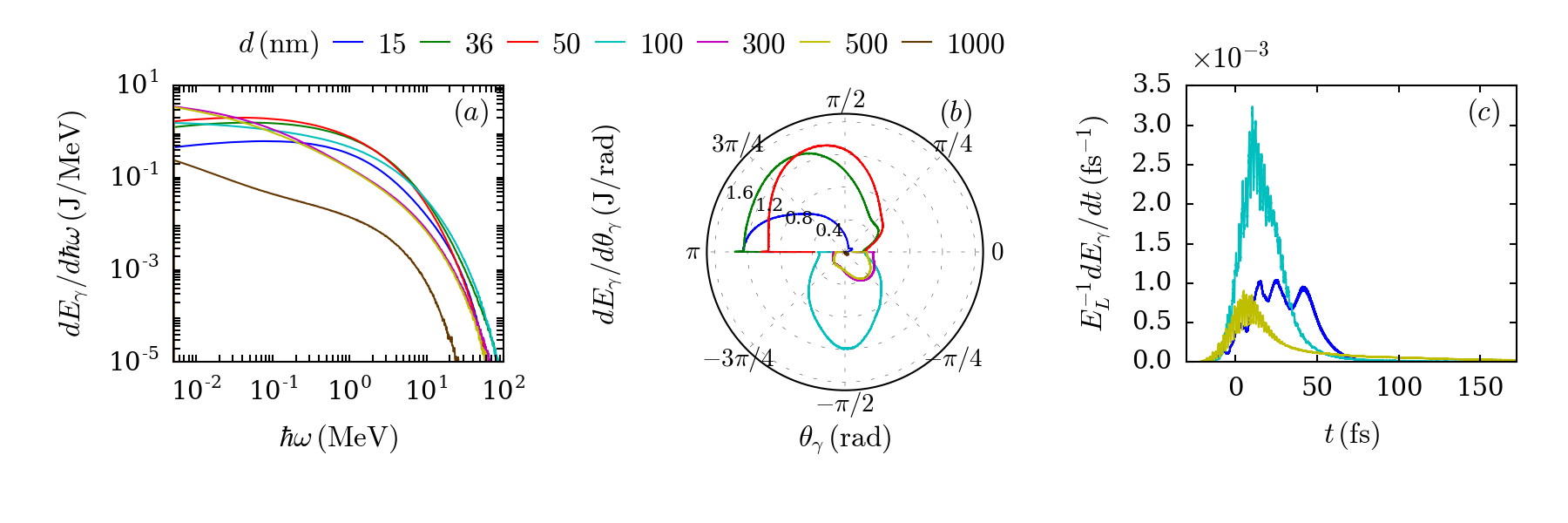}
\caption{Variations of the synchrotron emission with the wire width $d$: (a) energy spectra, (b) angle-resolved radiated energy and
(c) time-resolved radiated power (normalized to the total laser energy $E_L$). Each color represents a different value of $d$ (in $\mathrm{nm}$
units) as indicated in the legend of panel (a). Angles in (b) are defined as $\theta_\gamma = \arccos(k_{\gamma,x}/k_\gamma) \in (0,\pi)$
and the resulting angular distribution is then symmetrized with respect to $\theta_\gamma = 0$. All plotted quantities are integrated
over the simulation domain.} 
\label{fig:figure9a_9c}
\end{figure}

We now set the interwire spacing to the value maximizing the radiation efficiency, $D=1\,\mu\mathrm{m}$, and vary the wire width in the
set of values $d \in (15,36, 50, 100, 300, 500, 1000)\,\mathrm{nm}$. Note that the value $d=1\,\mu\mathrm{m}$ corresponds to a uniform 
solid-density target. The resulting energy-angle photon spectra and radiation dynamics are displayed in Fig.~\ref{fig:figure9a_9c}(a-c).

%A key element to explain the radiation pattern is to first understand the laser propagation depending on the wire diameter $d$. We evaluated the laser 
%front velocity in the nanowire-array averaged on the transverse direction $y$. We noted that it is maintained in-depth and transmitted through the rear of 
%the array for $d\leq 50 \, \mathrm{nm}$ in the RSIT regime. The onset of hole boring regime occurs for $d\geq 100 \, \mathrm{nm}$. This information is 
%significant as the regimes of radiation described in the next paragraphs are closely linked to the laser propagation regime in the nanowire-array. It was 
%previously demonstrated that for a 1D linearly polarised laser ($a_{0}=10$ and plasma density $n_{e}=4n_{c}$), a temporal transition from a RSIT dominated 
%regime to a hole boring one takes place \cite{NJPWeng2012} after $100$ laser periods. We do not observe such a behavior for our cases. We charge this 
%mainly on the fact we employ short ($10$ laser period) and time dependent pulses instead of uniform and infinite ones \cite{NJPWeng2012}.

For $d \lesssim \delta_{acc} \simeq 30\,\mathrm{nm}$, most of the electrons are expelled from the wires by the laser field, hence
leading to fast (\emph{i.e}, before ion expansion) homogenization of the plasma profile at the average density $n_\mathrm{av}=n_e d/D$.
For $d=15\,\mathrm{nm}$, one has $n_\mathrm{av} \simeq 7n_c$, which falls into the regime of relativistic self-induced transparency (RSIT).
Such plasma conditions have been shown to favor the RESE process \cite{PRLBrady2012}: the electrons, pushed by the ponderomotive force at
the laser front, are periodically reinjected back into the laser wave by the charge separation field. Their momentum ($\sim a_0 m_ec$) then
forms an angle of $\sim \pi$ with the laser wavevector, which maximizes the quantum parameter $\chi_e \sim 2 a_0^2 cB_0/E_S \sim 5\times 10^{-6} a_0^2$
and the subsequent synchrotron radiation in the backward direction. Figure~\ref{fig:figure9a_9c}(b) confirms this prediction, showing that
practically all the radiation is then directed backwards. In Fig.~\ref{fig:figure9a_9c}(c), we observe a temporal modulation of the radiated
power at a period of $\sim 15\,\mathrm{fs}$, of the same order as the theoretical estimate $\tau_\mathrm{RESE}=a_0/(n_e \omega_0)\simeq 7\,\mathrm{fs}$
derived for RESE in uniform plasmas \cite{PRLBrady2012}. This period is significantly larger than that of the $2\omega_0$ oscillations
arising in the SDE regime (see the curve with $d=500\,\mathrm{nm}$, corresponding to $n_\mathrm{av}=240n_c$).

As discussed below [see Fig.~\ref{fig:figure15a_15c}(a) in Sec.~\ref{sec:comp_unif_targets}], we have checked the occurrence of RSIT by measuring the
effective propagation velocity of the laser front in the homogenized plasma, in similar fashion to Ref.~\cite{NJPWeng2012}. RSIT is found to occur
for $d \lesssim 50-100\,\mathrm{nm}$, thus leading to significant laser transmission across the plasma. For wire widths
$\gtrsim 100\,\mathrm{nm}$, the homogenized plasma becomes opaque to the laser light, which then propagates at a much reduced speed through
hole boring (HB) \cite{NJPWeng2012}.

The synchrotron spectra of Fig.~\ref{fig:figure9a_9c}(a) show that the maximum photon energy weakly varies ($\hbar \omega_\mathrm{max}\simeq 50-70\,\mathrm{MeV}$),
and in a non-monotonic way, for $15\le d \le 300\,\mathrm{nm}$. The most notable variation occurs when the wire width is increased from $d=300\,\mathrm{nm}$
to $d=500\,\mathrm{nm}$, leading to $\hbar \omega_\mathrm{max}$ decreasing from $\sim 50\,\mathrm{MeV}$ to $\sim 20\,\mathrm{MeV}$. More interestingly, it
is found that the mean photon energy is maximized in the RSIT regime: for $d=15\,\mathrm{nm}$, we obtain $\langle \hbar\omega \rangle \simeq 0.45\,\mathrm{MeV}$,
much higher than for $d \geq 300\,\mathrm{nm}$, which leads to a relativistically overdense homogenized plasma ($n_{av} = 144n_c$) and
$\langle \hbar\omega \rangle \simeq 0.14 \,\mathrm{MeV}$.

\begin{figure}
\centering
\includegraphics[scale=0.8]{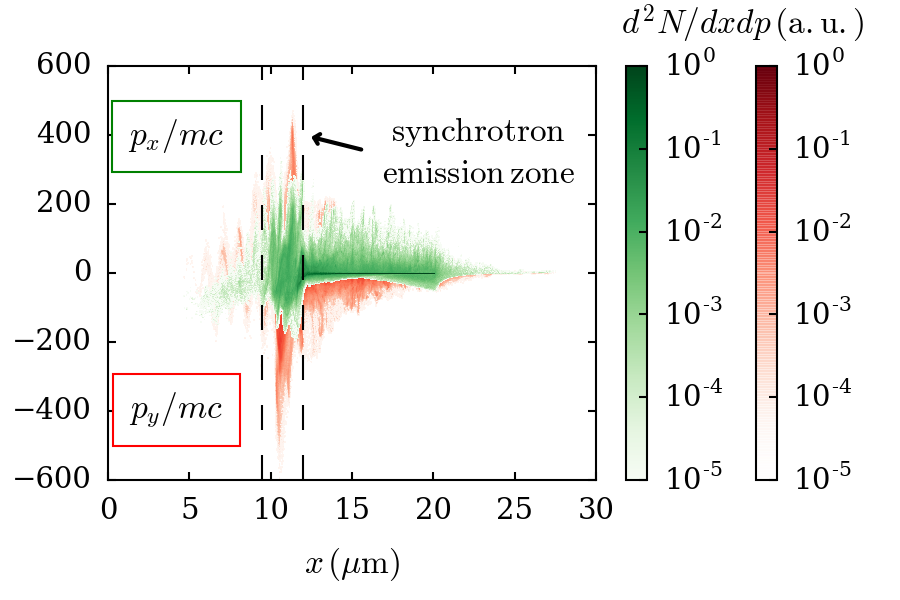}
\caption{Electron $x-p_x$ (green colormap) and $x-p_y$ (red colormap) phase spaces at $t=+40\,\mathrm{fs}$. The nanowire-array parameters
are $d=100\,\mathrm{nm}$, $D=1\,\mu\mathrm{m}$ and $L=10\,\mu\mathrm{m}$, giving rise to transversally oscillating electron synchrotron emission
(TOEE).}
\label{fig:figure10}
\end{figure}

The case of $d=100\,\mathrm{nm}$, close to the RSIT/HB threshold, yields the highest maximum photon energies [Fig.~\ref{fig:figure9a_9c}(a)] but also,
and more significantly,  to a radiated energy concentrated in the transverse direction, $\theta_\gamma = \pi /2$ [Fig.~\ref{fig:figure9a_9c}(b)].
This  particular radiation  pattern corresponds to the TOEE regime evidenced in Ref.~\cite{PoPChang2017}. In this mechanism, a balance is established
between the laser ponderomotive force and the charge-separation field at the irradiated plasma front. This causes the electrons to predominantly
oscillate in the transverse plane, thus inducing a mainly transverse synchrotron emission. This particular electron dynamics stands out in
Fig.~\ref{fig:figure10}, which superimposes the $x-p_x$ (green colormap) and $x-p_y$ (red colormap) electron phase spaces at $t=+40\,\mathrm{fs}$. 
Around the front side of the target where most of the radiation is emitted, the electron distribution is clearly more extended in the transverse
direction than in the longitudinal direction.
% As a consequence, the electrons are pushed and pulled several times such that their longitudinal momentum becomes comparable with, or even
% lower than its transverse momentum. This mechanism has already been evidenced with test particles \cite{PoPChang2017}. We therefore choose
% to study how much it is affected by Radiation Reaction through the observation of the phase space of electrons in Fig.~8 with and without the
% activation of the synchrotron module.  We note that the longitudinal and the transverse amplitude of the electron distribution
% is reduced when RR is taken into account. This can be explained as the most energetic electrons are in a semi-classical regime of emission where
% they lose a sizeable fraction of their kinetic energy. Indeed their quantum parameter can be approximated by $\chi \simeq \gamma a_0 / E_s \simeq 0.06$.
% The difference between the two curves confirms that most of the radiation comes from electrons with a large transverse momentum rather than a longitudinal one.
As the wire width is decreased (resp. increased) from $d \simeq 100\,\mathrm{nm}$, the radiation pattern is shifted to the backward (resp.
forward) direction, characteristic of the RESE (resp. SDE) mechanism.

\begin{figure}
\centering
\includegraphics[scale=1.]{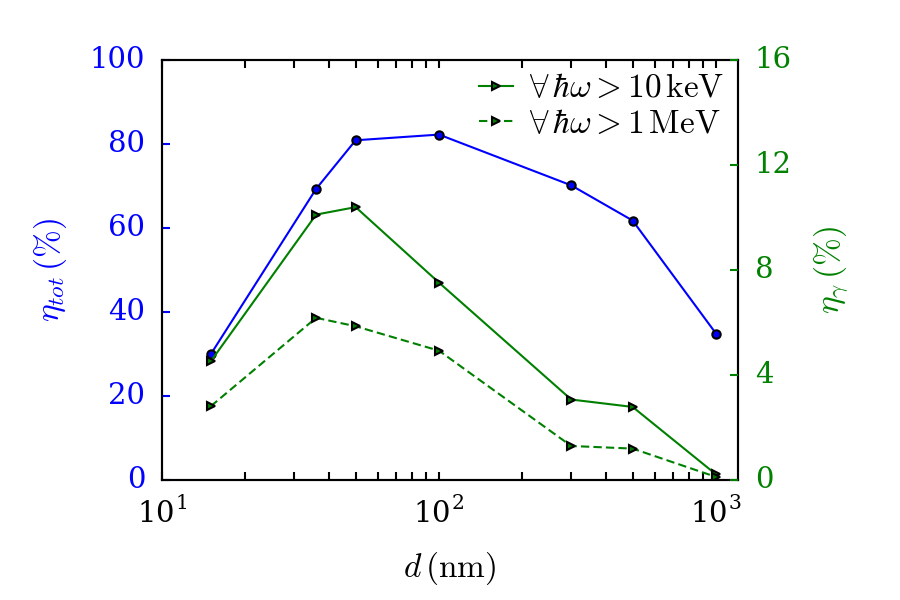}
\caption{Variations with the wire width ($d$) of the total absorbed laser energy fraction ($\eta_{tot}$, blue circles) and radiation
conversion efficiency ($\eta_\gamma$, green triangles). The radiation conversion efficiency is computed for two photon energy thresholds: 
$\hbar \omega \geq 10\,\mathrm{keV}$ (green solid) and $\hbar \omega \geq 1\,\mathrm{MeV}$ (green dashed). All quantities are integrated
over the whole simulation duration.}
\label{fig:figure11}
\end{figure}

Figure~\ref{fig:figure11} displays the wire-width dependence of the total laser absorption ($\eta_{tot}$) and radiation conversion efficiencies
($\eta_\gamma$) into $>10\,\mathrm{keV}$ and $>1\,\mathrm{MeV}$ energy photons.  The laser absorption rises from $\eta_{tot} \sim 30\,\%$
at $d=15\,\mathrm{nm}$ to a maximum of $\sim 80\,\%$ at $d=50-100\,\mathrm{nm}$, before dropping to  $\sim 35\,\%$ in the uniform-density
case ($d=1\,\mu\mathrm{m}$). While the increase in $\eta_{tot}$ at low wire widths is accompanied by similar rises in the $\eta_\gamma$ 
curves, the latter attain their maxima (at $d \simeq 36-50\,\mathrm{nm}$) slightly before $\eta_{tot}$. A peak value of $\eta_\gamma \sim 10.4\,\%$
(resp. $6.1\,\%$) for $\hbar \omega \ge 10\,\mathrm{keV}$ (resp. $>1\,\mathrm{MeV}$) is obtained at $d=50\,\mathrm{nm}$ (resp. $d=36\,\mathrm{nm}$).
Moreover, the two $\eta_\gamma$ curves show a faster decrease at large $d$ than $\eta_{tot}$.  To quantify this, let us compare the cases of 
$d=36\,\mathrm{nm}$ and $d=300\,\mathrm{nm}$: although both widths give rise to similar absorption fractions ($\eta \simeq 70\,\%$), the photon
yield at $d=36\,\mathrm{nm}$ is $\sim 3$ times larger than at $d=300\,\mathrm{nm}$. This marked difference follows from the distinct plasmas
produced by the electron-depleted exploding wires: at $d=36 \,\mathrm{nm}$, a relativistically undercritical plasma ($n_\mathrm{av} = 17n_c$)
forms, which triggers a RESE-type mechanism more efficient than SDE that arises in the overcritical plasma ($n_{av} = 144n_c$)
generated at $d=300 \,\mathrm{nm}$. Finally, we note that at $d=15\,\mathrm{nm}$, a sizable fraction ($\sim 70\,\%$) of the laser energy is transmitted 
across the array, which mechanically reduces the radiated energy fraction.

%In Ref.~\cite{PoPLecz2017}, the synchrotron radiation in a nanowire-array target was studied for diameters in the range $10\le d \le 90\,\mathrm{nm}$
%and an electron density of $7\times 10^{22}\,\mathrm{cm}^{-3}$ inside the wires. After a few laser cycles, the wires are depleted of all electrons and form
%a density favorable to Direct Laser Acceleration (DLA) of the electrons and, therefore, to strong betatron radiation. An optimum is reached for a diameter of
%$60\,\mathrm{nm}$ corresponding to an average electron density $n_{av} = 0.1n_c$. We stress that this optimized diameter is obtained in another regime
%that the one explored in our study. The optimum we obtain is at $d=50\,\mathrm{nm}$, leading to an average density $n_{av}=24 n_c$. DLA is
%therefore mitigated as the electrons cannot stay in phase with the laser field at such high densities.

%The above parametric scan on the wire width has confirmed the findings of the previous scan on the wire spacing: at small $d$, the fast homogenization
%of the expanding wires into a relativistically transparent plasma favorable to backward-directed RESE \cite{PRLBrady2012}. At large $d$, the homogenized
%electron density is relativistically overcritical, and induces a weaker synchrotron efficiency directed in a large forward cone angle.

% section 2 - subsection 3 : atomic number dependence
\subsection{Changing the ion mass and the laser intensity}

We now demonstrate that modifying other key parameters of the interaction such as the wire material or the laser intensity can also
enable switching between the previously discussed radiation mechanisms. To this goal, we first replace, in the most efficient
configuration for $\gamma$-ray production ($D=1\,\mu\mathrm{m}$, $d=36\,\mathrm{nm}$), the neutral carbon atoms ($Z=6$) by either
copper ions ($Z=29$) with a $5+$ initial ionization degree and a solid density $n_{Cu}=80 n_c$, or gold ions ($Z=79$) with a $14+$
initial ionization degree and a solid density $n_{Au}=55 n_c$. Second, we vary the laser intensity in the range $I=10^{21}-10^{23}\,\mathrm{Wcm}^{-2}$
for two values of the wire widths: $d=100\,\mathrm{nm}$ and $d=300\,\mathrm{nm}$.
% We finally give the scaling of the radiative conversion efficiency $\eta_{\gamma}$ and relate it to the one in the relativistically transparent regime
% \cite{PRLBrady2012} and the overdense plasma case \cite{EPJSJi2014}.

\begin{figure}
\centering
\includegraphics[scale=0.8]{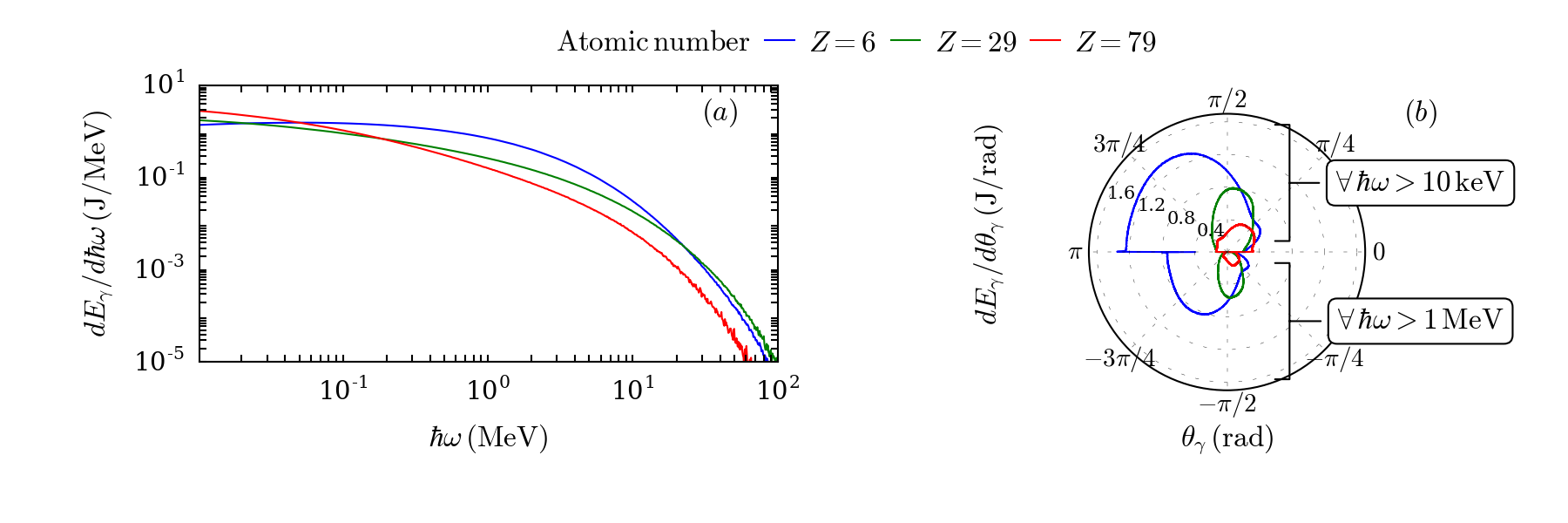}
\caption{Wire-material dependence of the (a) energy-resolved and (b) angle-resolved radiated energy. The blue, green and red curves
correspond, respectively, to C, Cu and Au wires. The top (resp. bottom) half of panel (b) is associated with a photon energy threshold
of $10\,\mathrm{keV}$ (resp. $1\,\mathrm{MeV}$). All spectra are integrated over the simulation duration. Angles in (b) are defined as
$\theta_\gamma = \arccos(k_{\gamma,x}/k_\gamma) \in (0,\pi)$, and the resulting angular distribution is symmetrized with respect to
$\theta_\gamma=0$.}
\label{fig:figure12a_12b}
\end{figure}

The energy-resolved radiated energy displayed in Fig.~\ref{fig:figure12a_12b}(a) indicates that the average photon energy is decreased by the use of
copper  ($0.27\,\mathrm{MeV}$) and gold ($0.14\,\mathrm{MeV}$) compared to carbon ($0.41\,\mathrm{MeV}$). The radiation efficiency above $10\,\mathrm{keV}$ also drops with increasing atomic number (from $\sim 10.1\,\%$ in carbon to $4.6\,\%$ in copper and
$2.9\,\%$ in gold), in spite of a slightly enhanced laser absorption in copper and gold ($\eta_{tot} \sim 80\,\%$) than in carbon ($\sim 70\,\%$,
see Fig.~\ref{fig:figure11}). In light of our previous results, the reason for this difference is that the homogenized electron density ($n_{av}=17n_c$)
in the carbon wires lies in the RSIT regime, prone to RESE. In contrast, the copper (resp. gold) wires produce a higher-density plasma, $n_{av} =80n_c$
(resp. $n_{av} =3000n_c$), opaque to the laser field, which favours TOEE (resp. SDE). This transition from RESE to SDE through TOEE is supported by
the angular radiation patterns shown in Fig.~\ref{fig:figure12a_12b}(b): both for the $10\,\mathrm{keV}$ and $1\,\mathrm{MeV}$ photon energy thresholds,
we clearly see that the emission evolves from a mainly backward radiation in the carbon target to a predominatly transverse radiation in copper and
to a forward directed radiation in gold.

In the gold case, we observe ionization rates up to $Z^*=70$ at the laser-target interface, consistent with Ref.~\cite{SABargsten2017} where a similar
setup is numerically considered. Also, the synchrotron photon yield above $1\,\mathrm{MeV}$ ($\sim 1.9\,\%$) is about $60\,\%$ of the yield
above $10\,\mathrm{keV}$, similarly to the carbon and copper targets. Although this performance is not optimal due to too dense a homogenized plasma, it can be put in
perspective with the record $\sim 20\,\%$ conversion efficiency into $>1\,\mathrm{keV}$ photons which has been recently reported using gold nanowires
driven by a $4\times10^{19}\,\mathrm{Wcm}^{-2}$, $55\,\rm fs$ laser pulse \cite{OHollinger2017}. Rather than synchrotron emission, x-ray radiation
in this experiment is caused by atomic physics processes (atomic line emissions, photorecombination and Bremsstrahlung). Another difference with our 
study is that, due to lower laser intensity, and hence slower nanowire expansion, the highest x-ray yield is found for significantly smaller interspacings 
($\sim 0.1\,\mu\mathrm{m}$). The measured x-ray yield, however, rapidly drops with increasing photon energies (below $1\,\%$ for $\hbar \omega > 6\,\mathrm{keV}$).
These results should stimulate further theoretical work on the radiation efficiencies of atomic physics and synchrotron processes as functions of the
laser and nanowire parameters.

%They were shown to be maximized by enabling a large radiative to hydrodynamic energy loss rate ratio. For higher photon energies this record conversion
%efficiency drops to $\simeq 1\,\%$ for $\hbar \omega >6 \, \mathrm{keV}$. Despite the two basic mechanisms at stake are fundamentally different,
%they are both influenced by the transverse expansion of the wires.

%The main point we can extract from this analysis is that to optimize a synchrotron source from a nanowire-array target, one has to first increase
% the spatio-temporal overlap of electrons and high field zones and second the laser absorption. This argument is complementary to the observations
% in Figures \ref{fig:figure10} and %\ref{fig:figure11} where we corroborate the well-known fact that laser absorption is a necessary, but not sufficient, 
% step toward getting an efficient high energy photon emission. 

\begin{figure}
\centering
\includegraphics[scale=0.8]{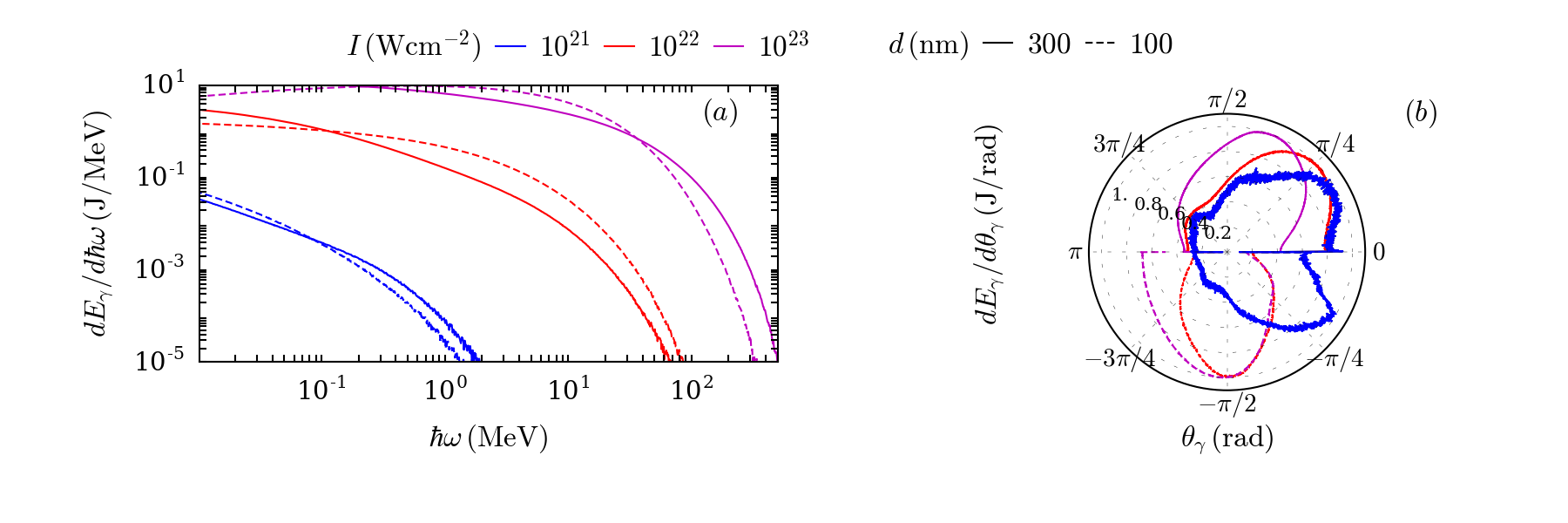}
\caption{Laser-intensity dependence of (a) the energy-resolved and (b) angle-resolved radiated energy (above $10\,\mathrm{keV}$).
Each color stands for a particular value of $I$ as indicated in the legend. The solid (resp. dashed) curves correspond to a wire width
$d=300\,\mathrm{nm}$ (resp. $100\,\mathrm{nm}$). The interwire spacing is set to $D=1\,\mu\mathrm{m}$. Angles in (b) are defined
as $\theta_\gamma = \arccos(k_{\gamma,x}/k_\gamma) \in (0,\pi)$, and the resulting angular distribution is symmetrized with respect
to $\theta_\gamma=0$.}
\label{fig:figure13a_13b}
\end{figure}

We now return to carbon nanowires and examine the photon distributions produced in the laser intensity range $10^{21}\le I \le 10^{23}\,\mathrm{Wcm}^{-2}$.
Figure~\ref{fig:figure13a_13b}(a) reveals that the photon generation at $I=10^{21}\,\mathrm{Wcm}^{-2}$ occurs with the same efficiency for the two chosen
values of the wire width, $d=100\,\mathrm{nm}$ (dashed lines $\eta_\gamma \simeq 0.09\,\%$) and $d=300\,\mathrm{nm}$ (solid lines $\eta_\gamma \simeq 0.08\,\%$).
The case of $d=300\,\mathrm{nm}$, however, leads to higher maximum ($\hbar\omega_\mathrm{max} =1.4\rightarrow 1.8\,\mathrm{MeV}$) and average
($\langle \hbar \omega \rangle = 32\rightarrow 42\,\mathrm{keV}$) photon energies.
%A possible explanation for this behavior is the local field amplification ($B_z /B_{0}=2.7a_0$) that is seen to take place around the thicker wires.
At higher intensity ($I\geq 10^{22}\,\mathrm{Wcm}^{-2}$), by contrast, the average photon energy is much larger at $d=100\,\mathrm{nm}$
than at $d=300\,\mathrm{nm}$ ($400 \, \mathrm{keV}$ \emph{vs.} $140\,\mathrm{keV}$). 
This stems from the fact that the expanded plasma then becomes relativistically transparent, whereas it remains opaque at $d=300\,\mathrm{nm}$
(even at $I=10^{23}\,\mathrm{Wcm}^{-2})$. Furthermore, the fraction of laser energy converted into $\ge 10\,\mathrm{keV}$ photons
is always higher at $d=100\,\mathrm{nm}$ whatever the laser intensity in the studied range.

In Fig.~\ref{fig:figure13a_13b}(b), it is seen that the emission cone angle increases with increasing laser intensity. While at $d=300\,\mathrm{nm}$
the radiation remains forward-directed up to $I=10^{23}\,\mathrm{Wcm}^{-2}$, at $d=100\,\mathrm{nm}$ it is forward directed at $I=10^{21}\,\mathrm{Wcm}^{-2}$,
becomes concentrated in the transverse direction at $I=10^{22}\,\mathrm{Wcm}^{-2}$, and is mainly confined within angles $\ge \pi/2$ at $I=10^{23}\,\mathrm{Wcm}^{-2}$. 
Once again we stress that this evolution from SDE to RESE results from the onset of RSIT at high enough laser intensity.
In the latter case, the radiation is mostly carried by $\gamma$-ray photons: the radiation conversion efficiency above $1\,\mathrm{MeV}$ indeed reaches $\sim 43\,\%$,
hardly lower than the $\sim 47\,\%$ conversion fraction in $\ge 10\,\mathrm{keV}$ photons. 

\begin{figure}
\centering
\includegraphics[scale=0.8]{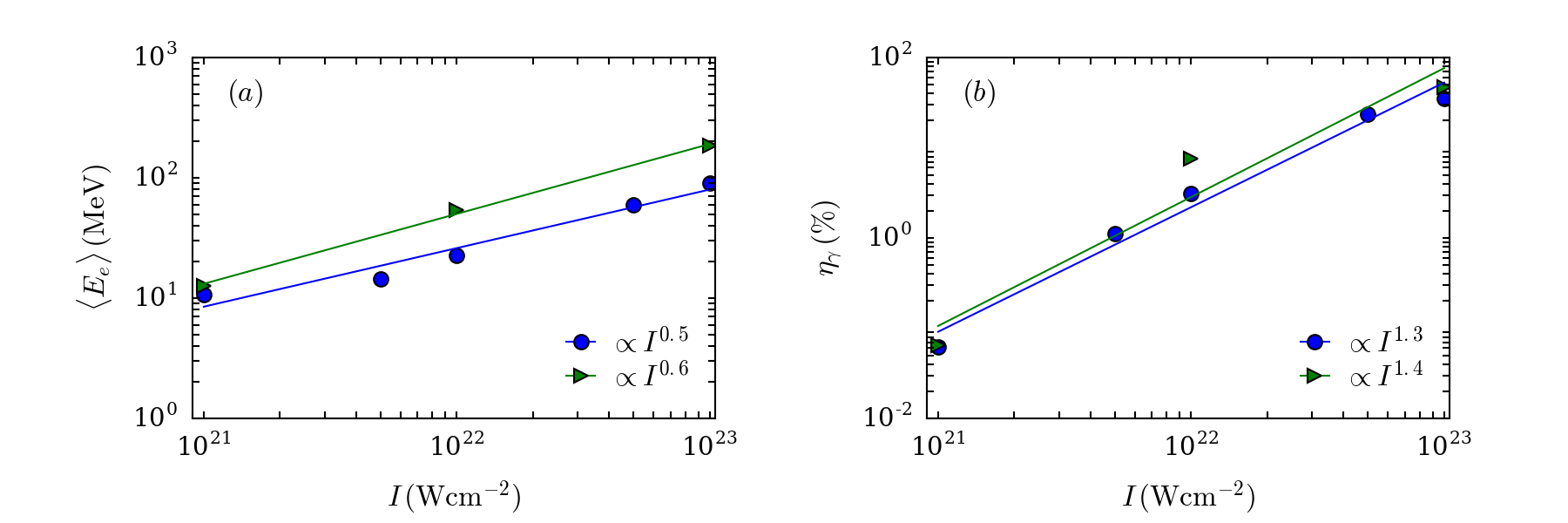}
\caption{(a) Hot-electron ($>0.511\,\mathrm{MeV}$) mean energy and (b) radiation conversion efficiency into $>10\,\mathrm{keV}$ photons
as a function of the laser intensity. The green triangles (resp. blue circles) correspond to a wire width $d=100\,\mathrm{nm}$ (resp. $300\,\mathrm{nm}$).
The interwire spacing is set to $D=1\,\mu\mathrm{m}$.}
\label{fig:figure14a_14b}
\end{figure}

Since the radiation power should scale approximately as $\chi_e^2$, it is worthwhile to inspect the variations of the mean hot-electron
energy $\langle E_e \rangle$ (counting all electrons above $0.511\,\mathrm{MeV}$) as a function of the laser intensity.
Figure~\ref{fig:figure14a_14b}(a) plots $\langle E_e \rangle$ for the wire widths $d=100\,\mathrm{nm}$ (green triangles) and $d=300\,\mathrm{nm}$
(blue circles). Both curves are consistent with an approximate scaling $\langle E_e \rangle \propto I^{0.5-0.6}$, quite close to the
fit $\langle E_e \rangle \propto I^{0.4}$, reported at lower intensities ($10^{18}\le I \le 3\times 10^{20}\,\mathrm{Wcm}^{-2}$) in
Ref.~\cite{PoPCao2010b}. The fact that $\langle E_e \rangle$ roughly obeys the well-known ponderomotive law \cite{PRLWilks1992} is not
\emph{a priori} obvious given the various heating mechanims possibly at play in the nanowire array: from Brunel-type acceleration at
the wire walls to stochastic heating in the interference field pattern inside the vacuum gaps, and ponderomotive acceleration in the
homogenized plasma. It should be noted that the acceleration of super-ponderomotive electrons was recently demonstrated in
the case of a nanowire array with $d=1.5\,\mu\mathrm{m}$ and $D=7\,\mu\mathrm{m}$, driven at $I\simeq 10^{21}\,\mathrm{Wcm}^{-2}$
\cite{PRLJiang2016}. The main difference between this work and ours is the plasma-filling time of the interstices: the large
interspacing in the experiment allows the laser to efficiently propagate between the wires, and energize electrons via the so-called
direct laser acceleration mechanism \cite{PRLJiang2016}. In our case, such an efficient laser penetration is hampered by the fast homogenization
of the nanostructure front, due to the comparatively lower interspacing $D=1\,\mu\mathrm{m}$ investigated (at $I=10^{21}\,\mathrm{W/cm}^{2}$,
the fraction of energy transmitted across the target is $<1\,\%$).

In Fig.~\ref{fig:figure14a_14b}(b) is plotted the radiation conversion efficiency (counting all photons above $10\,\mathrm{keV}$)
as a function of the laser intensity. The results can be approximately fitted to $\eta_\gamma \propto I^{1.3-1.4}$. This scaling
happens to fall in between the one found at undercritical densities in the RESE regime, $\eta_\gamma \propto I$ \cite{PRLBrady2012},
and the one observed at overcritical densities in the SDE regime, $\eta_\gamma \propto I^{3/2}$ \cite{EPJSJi2014}. This behavior
could be expected since both radiation regimes can arise in our broad intensity range. Regarding the radiation efficiency, these
two mechanisms mainly differ in the typical number of radiating electrons ($N_e$). In the underdense plasma, this number is
proportional to the areal density crossed by the laser, $N_e \propto n_e$; in an overdense plasma, this number scales as the
areal density of the compressed electron layer at the target front, $N_e \propto I^{1/2}$. Since $\eta_\gamma \propto N_e \chi_e^2/I$
and $\chi_e \propto I$, we thus expect $\eta_\gamma \propto I$ for RESE and $\eta_\gamma \propto I^{3/2}$ for SDE.   

%In the two descriptions, the radiated energy for one electron, denoted by $E_{\gamma}$ is assumed to be proportional to the classical
%radiated power $P_{cl} \propto \chi_{e}^{2}\propto I^{2}$ under the assumption that the energetic electrons temperature follows the
%ponderomotive scaling. We underline that in our case this hypothesis is verified in Fig.~\ref{fig:figure14a_14b} (a). By means of
%different arguments they both determine that the number of electrons, denoted by $N_e$, satisfies $N_e\propto I^{1/2}$. The difference
%between the two regimes relies in the fact that the RESE emission occurs in successive breakdown events during which electrons are
%re-accelerated back by the charge separation field they generated at the front of the target. These events which period scales as
%$I^{-1/2}$ results therefore in a radiative conversion efficiency that scales linearly with the intensity whereas it scales as $I^{3/2}$
%in the overdense regime.

% section 2 - subsection 4 : comparison with uniform targets
\subsection{Comparison with uniform-density targets}
\label{sec:comp_unif_targets}

The dominant radiation processes that we have highlighted in nanowire arrays appear similar to those identified in previous simulation
studies considering uniform plasmas. This is so because, under the present interaction conditions, the nanostructure is largely smoothed
out during the laser pulse, so that a large part of it experiences a significantly homogenized plasma. One may then question the advantage,
regarding synchrotron radiation, of using nanowire arrays compared to uniform plasmas at sub-solid densities. To answer this question, we
have conducted a set of simulations considering a $10\,\mu\mathrm{m}$-thick carbon layer of uniform (free electron) density varying from
$Zn_i = 7n_c$ to $480 n_c$ (solid density). This density range corresponds to that achieved in fully homogenized nanowire arrays
($n_{av}=Zn_C d/D$) when increasing the wire width from $d=15\,\mathrm{nm}$ to $1\,\mu\mathrm{m}$ at fixed spacing $D=1\,\mu\mathrm{m}$.
The laser intensity is set to  $I=10^{22}\,\mathrm{Wcm}^{-2}$. 

\begin{figure}
\centering
\includegraphics[scale=0.8]{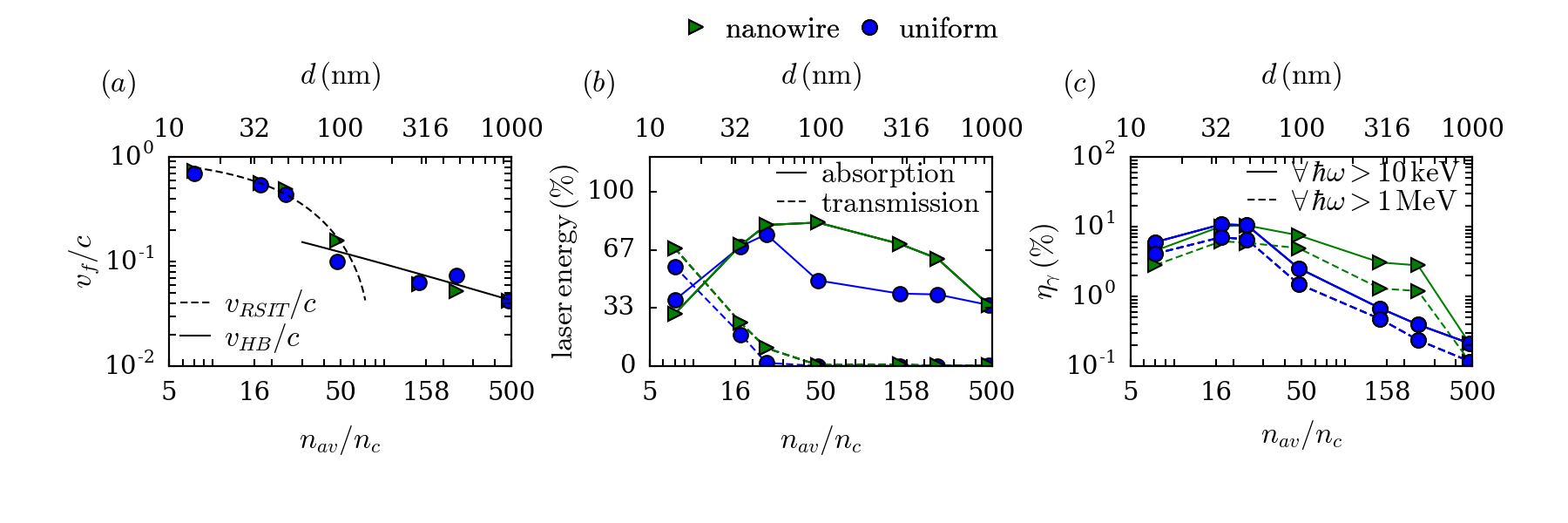}
\caption{Comparison between carbon nanowire arrays (green triangles) and uniform-density targets (blue circles): (a) propagation velocity of the
laser front; (b) total laser absorption (solid lines) and transmission (dashed lines); (c) radiative conversion efficiencies into $>10\,\mathrm{keV}$
(solid lines) and $>1\,\mathrm{MeV}$ photons (dashed lines). Results from nanowire-array (resp. uniform-density) targets are plotted as functions
of the wire width $d$ (resp. the average electron density $n_{av}$). In (a), the black solid and dashed lines plot the theoretical front velocities
$v_{RSIT}$ and $v_{HB}$, respectively (see text). In (b) and (c), all quantities are integrated over the simulation duration.}
\label{fig:figure15a_15c}
\end{figure}

First, we examine the transition between plasma transparency (RSIT) and opacity (HB), which appears critical in determining the properties of
the synchrotron emission. To properly identify the regime of laser-plasma interaction, we have tracked the position of the laser front in the
target, $x_f(t)$, defined such that $a\left(x_f(t),t \right)=\max_x a(x,t)/2$, with $a(x,t)$ being the $y$-averaged dimensionless laser field.
This definition is similar to that used in Ref.~\cite{NJPWeng2012} except that, due to our short pulse duration, we use $\max_x a(x,t)/2$
instead of $a_0/2$ as is relevant to a semi-infinite pulse. For each simulation, $v_f$ is evaluated from a linear regression fit of $x_f(t)$.
Figure~\ref{fig:figure15a_15c}(a) plots $v_f$ as a function of the wire width ($d$) in the nanowire-array case, and of the electron density
($n_e \equiv n_{av}$) in the uniform-plasma case. Both target types lead to a similarly decreasing curve for $v_f$, which drops from
$v_f/c \simeq 0.7$ at $n_{av}=7n_c$ down to $v_f/c \simeq 0.2$ at $n_{av}=32 n_c$. This parameter range corresponds to RSIT, and we
have further checked that the laser wave then overlaps with the plasma electrons and ions, as expected \cite{NJPSiminos2017}. Nanowire
arrays tend to yield slightly faster laser propagation, which is ascribed to inhomogeneity effects. For completeness, we have plotted (as a black
solid line) the front velocity estimated in Ref.~\cite{NJPWeng2012} in a simpler setting (1D geometry, semi-infinite pulse, no synchrotron losses),
$v_{RSIT}/c \simeq \exp(-2n_{av}/n_{cr})\sqrt{1-n_{av}/n_{cr}}$, where $n_{cr}\simeq 0.89a_0 n_c$ in the ultrarelativistic regime. Despite the
short duration and time-varying intensity of our laser pulse, correct agreement is found between $v_f$ and $v_{RSIT}$ up to $n_{av} \simeq 48n_c$
(or $d \simeq 100\,\mathrm{nm}$), where the transition from RSIT to HB occurs, also corresponding to the transition threshold between RESE
and SDE  [see Figs.~\ref{fig:figure9a_9c}(b) and \ref{fig:figure10}]. At higher $n_{av}$ or $d$, the front velocity approximately matches the
theoretical HB velocity (black dashed line), $v_{HB}/c \simeq \Pi/(1+\Pi)$, where $\Pi=\sqrt{I Z/A m_e n_{av} c^3}$ \cite{NJPWeng2012}. 

In Fig.~\ref{fig:figure15a_15c}(b) are plotted the absorbed and transmitted laser energy fractions as functions of the wire width ($d$) in
the nanowire-array case, and of the plasma density ($n_e \equiv n_{av}$) in the uniform-target case. Similarly, Fig.~\ref{fig:figure15a_15c}(c)
plots, for both target types, the variations with $n_{av}$ and $d$ of the conversion efficiencies into $>10\,\mathrm{keV}$ and  $>1\,\mathrm{MeV}$
photons. In uniform targets, the laser absorption strongly increases (from $\eta_{tot} \sim 35\,\%$ to $\sim 75\,\%$) with increasing density in
the range $7 \le n_{av} \le 24 n_c$. Similar variations are found in nanowire arrays with same equivalent density (\emph{i.e.},
$15\le d \le 50\,\mathrm{nm}$), with the differences, however, that $\eta_{tot}$ is a bit smaller ($\sim 30\,\%$) at $n_{av}=7n_c$, but larger
($\sim 80\,\%$) at $n_{av}=24n_c$. In this parameter range, the interaction takes place in the RSIT regime in both targets, yet the transmitted
laser fraction is always a bit larger in nanowire arrays, reaching $\sim 70\,\%$ at $n_{av}=7n_c$ and $\sim 10\,\%$ at $n_{av}=24n_c$). The most
pronounced difference between the two target types arises at larger $n_{av}$ or $d$: while the laser absorption in uniform targets abruptly drops
beyond $n_{av}=24n_c$, (down to $\eta_{tot} \simeq 45\,\%$ at $n_{av}=64n_c$, and $\eta_{tot} \simeq 35\,\%$ at solid density), it stays at a
high level ($\gtrsim 70\,\%$) up to $n_{av}=144n_c$ (\emph{i.e.}, $d=300\,\mathrm{nm}$).

The general trends observed for the laser absorption also hold for the synchrotron radiation. While uniform targets yield slightly better
radiation efficiencies at $n_{av}=7n_c$ ($\eta_\gamma \simeq 6\,\%$ \emph{vs.} $\sim 4.5\,\%$, for $\hbar \omega > 10\,\mathrm{keV}$),
both setups give very similar maximum efficiencies, $\eta_\gamma \simeq 10\,\%$ (resp. $\sim 6\,\%)$ for $\hbar \omega > 10\,\mathrm{keV}$
(resp. $>1\,\mathrm{MeV}$) in the range $n_{av}=17-24n_c$ (\emph{i.e.}, $d=36-50\,\mathrm{nm}$). The robustness of the laser absorption
enhancement in nanowire arrays is accompanied by a similar robustness of the radiation efficiency, which remains relatively high,
$\eta_\gamma>3\,\%$ (resp. $>1\,\%$) for $\hbar \omega > 10\,\mathrm{keV}$ (resp. $>1\,\mathrm{MeV}$) up to $n_{av} = 240n_c$
($d=500\,\mathrm{nm}$). By contrast, the radiation yield from uniform targets decreases rapidly after its maximum: $\eta_\gamma$ drops
by a factor $\sim 2.5$ when $n_{av}$ is increased from $24n_c$ to $32n_c$, and falls below $3\,\%$ for $n_{av} \ge 64n_c$. 

If we restrict our analysis to the forward radiation ($\theta_\gamma\le 30^\circ$), we find that the highest yield into $>1\,\mathrm{MeV}$
photons ($\eta_\gamma \simeq 0.4\,\%$) is provided by a nanowire array of width $d=36\,\mathrm{nm}$, yet with little variation ($<10\,\%$)
in the $36\le d \le 100\,\mathrm{nm}$ range. Also, the highest yield into $>10\,\mathrm{keV}$ photons is observed for $d=300\,\mathrm{nm}$
($\eta_\gamma \simeq 0.7\,\%$), with $<10\,\%$ variation in the $36\le d \le 300\,\mathrm{nm}$ range. 

%% file: section3_LG.tex
%-----------------------------------------------------------------------
% section 3
%-----------------------------------------------------------------------
\section{\label{sec:3}Radiation enhancement by a reflective substrate}

We  now investigate whether a more realistic setup, whereby the nanowire array is coated on a solid-density
substrate, may substantially improve the synchrotron process. The rationale for this is that, for the parameters
(carbon wires with $D=1\,\mu\mathrm{m}$ and $d=36-100\,\mathrm{nm}$) previously found to yield the highest radiation
efficiencies ($\eta_\gamma \geq 8\,\%$), a sizable fraction of the laser energy (\emph{e.g.}, $\sim 25\,\%$ at
$d=36\,\mathrm{nm}$) shines through the target via RSIT. Making this transmitted light reflect off a plasma mirror
so as to interact with the hot electrons filling the nanowire array  could sustain the synchrotron emission, and
hence increase its efficiency.

\begin{figure}
\centering
\includegraphics[scale=0.8]{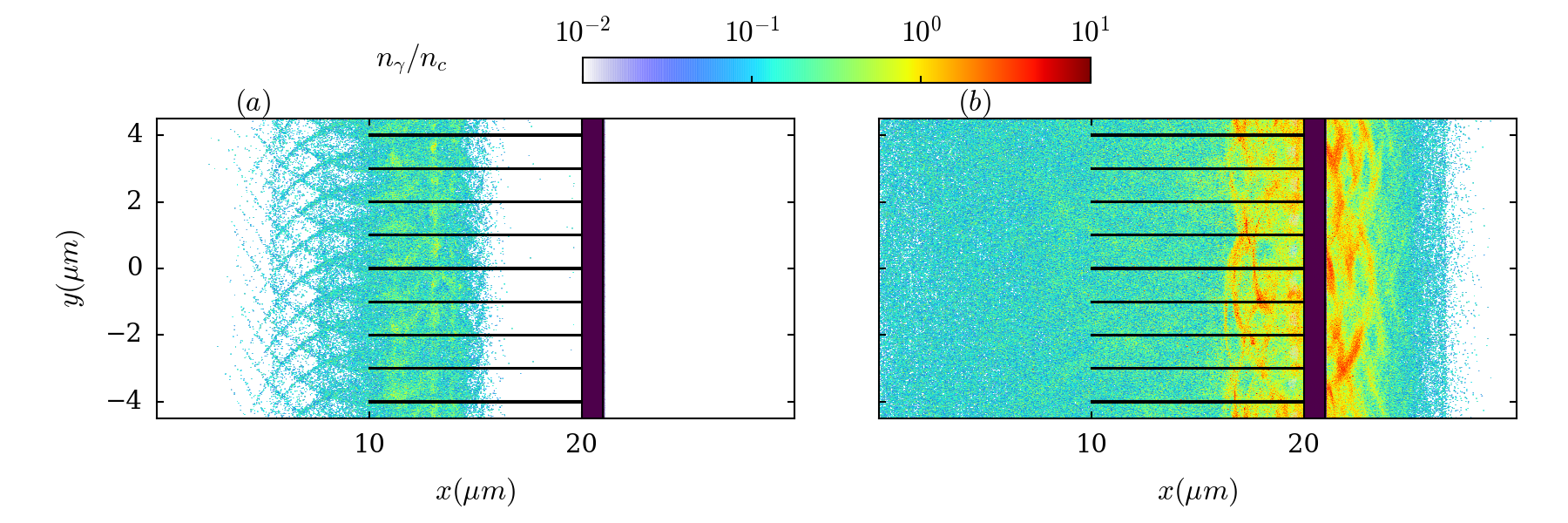}
\caption{Normalized density ($n_\gamma/n_c$) of the high-energy ($>1\,\mathrm{MeV}$) photons (a) before ($t=+13\,\mathrm{fs}$)
and (b) after the reflection ($t=+56\,\mathrm{fs}$) of the laser pulse off the copper foil at the target backside. The initial
target shape is shown in dark red.}
\label{fig:figure16a_16b}
\end{figure}
  
To test this scenario, we have performed a simulation in which a $1\,\mu\mathrm{m}$-thick copper foil is placed at the backside of a
carbon wire array with $D=1\,\mu\mathrm{m}$, $d=36\,\mathrm{nm}$ and $L=10\,\mu\mathrm{m}$. The Cu ions are initialized with $5+$
charge state and a density $n_{Cu}=80n_c$. As before, collisional and field ionizations are described. The laser pulse maximum
($10^{22}\,\mathrm{Wcm}^{-2}$) strikes the Cu foil at $t_r=+33\,\mathrm{fs}$.  For this simulation only, the $\gamma$-ray photons
($\hbar \omega \geq 1\,\mathrm{MeV}$) are advanced (ballistically) on the simulation domain. The evolution of their density is depicted
in Figs.~\ref{fig:figure16a_16b}(a,b). At $t=+13\,\mathrm{fs}<t_r$ [Fig.~\ref{fig:figure16a_16b}(a)], the wires have rapidly expanded
(in the leading edge of the laser) to form a relativistically underdense plasma ($n_e \simeq 17n_c$), in which synchrotron emission
occurs volumetrically mainly through RESE, as analyzed in Sec.~\ref{sec:variation_width}. At $t=+56\,\mathrm{fs} > t_r$
[Fig.~\ref{fig:figure16a_16b}(b)], high-density ($\sim 10n_c$) photon bunches are seen to radiate from the target backside.

\begin{figure}
\centering
\includegraphics[scale=0.8]{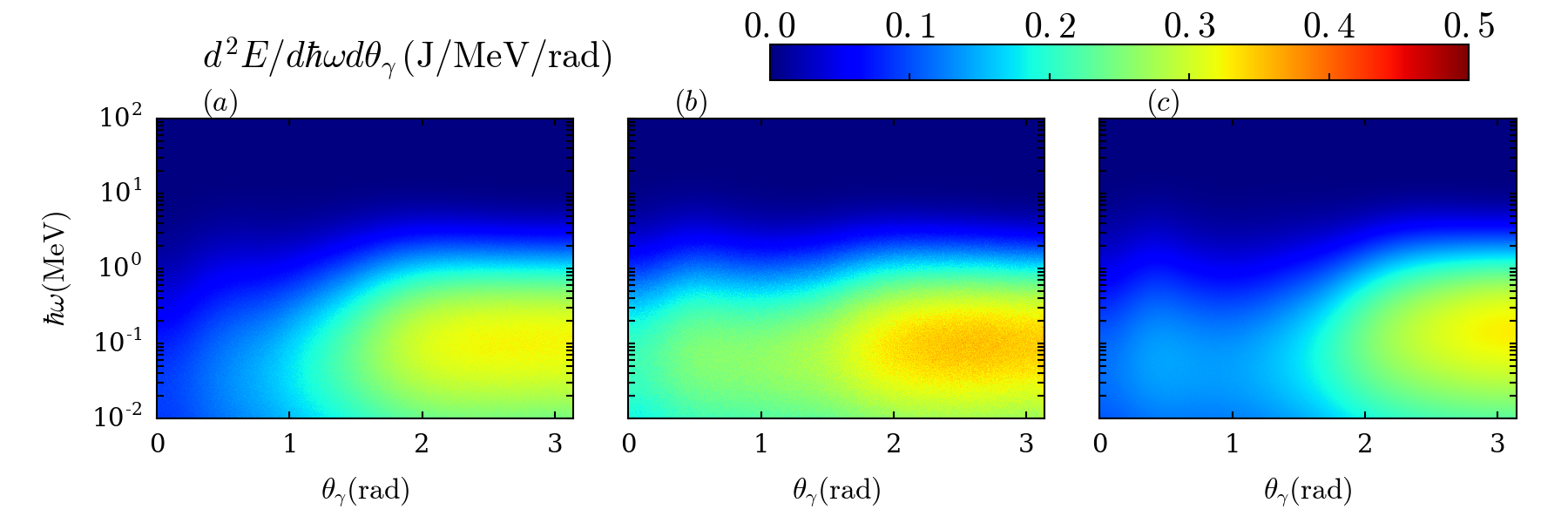}
\caption{Energy-angle spectrum of the radiated energy for (a) the optimized nanowire array ($d=36\,\mathrm{nm}$, $D=1\,\mu\mathrm{m}$)
without substrate, (b) the optimized nanowire array target with substrate, and (c) the optimized (and density-equivalent) uniform plasma
($n_e=16n_c$). Angles are defined by $\theta_\gamma = \arccos \left(k_{\gamma,x}/k_\gamma \right) \in(0,\pi)$. All spectra are integrated
over the simulation duration.} 
\label{fig:figure17a_17c}
\end{figure}

The resulting time-integrated energy-angle radiation spectrum is displayed in Fig.~\ref{fig:figure17a_17c}(b), and compared with that
obtained from the sole nanowire array [Fig.~\ref{fig:figure17a_17c}(a)]. Comparison of the two spectra reveals the generation of two
distinct photon groups. The first one originates from the interaction with the expanded wires, and is broadly distributed in the
backward direction ($\theta_\gamma =2-3\,\mathrm{rad}$) with mean energies $\sim 0.4\,\mathrm{MeV}$ (resp. $\sim 2.5\,\mathrm{MeV}$) for
$\hbar \omega >10\,\mathrm{keV}$ (resp. $>1\,\mathrm{MeV}$). The second one follows the reflection of the laser head off the foil,
and its interaction with the electrons still accelerated in the laser tail. As already stressed, the quantum parameter is maximized for the
forward-moving electrons that stream against the reflected pulse. Consequently, in this emission stage the radiated energy is mainly,
but not entirely, forward-directed, as seen by comparing Figs.~\ref{fig:figure17a_17c}(a) and (b). This secondary emission stage 
increases the integrated radiation efficiency to $\sim 13\,\%$ (\emph{vs.} $\sim 10\,\%$ without substrate, for $\hbar \omega >10\,\mathrm{keV}$).
Closer analysis reveals that out of the $\sim 26\,\%$ of laser energy hitting the Cu foil, approximately $13\,\%$ is further gained by
electrons and ions, $3\,\%$ is converted into photons, and $10\,\%$ escapes through the target front side.
%We underline that a significant (?) gain on the photon conversion efficiency is only possible if the incident and reflected parts of the laser 
%spatially overlap, which implies... This condition is met in our case}
% J'ai supprime cette phrase car la reflexion provoque toujours un chevauchement entre les parties incidente et reflechie sur laser, et ce, 
% quel que soit $L$. 

Finally, we show in Fig.~\ref{fig:figure17a_17c}(c) the energy-angle spectrum recorded from the optimized uniform-density target ($n_e=16n_c$),
giving a radiation efficiency $\eta_\gamma \sim 11\,\%$ into $> 10\,\mathrm{keV}$ photons. It corroborates our previous findings that optimized
nanowire arrays and uniform targets yield similar photon distributions. Notable differences, however, are visible: the backward-emission cone
angle is slightly narrower, and is complemented by a distinct, albeit weaker, forward emission around $\theta_\gamma \sim \pi/4 \,\mathrm{rad}$.

While the optimized nanowire array with substrate yields the highest radiation conversion efficiency, $\eta_\gamma=13\%$ (for $\hbar \omega >10\,\mathrm{keV}$),
its performance falls by an order of magnitude, as does that of the two other types, if we consider only photon energies $>1\,\mathrm{MeV}$ and
forward emission angles $\le 30\ensuremath{^\circ}$ (as would be relevant for, \emph{e.g.}, creating electron-positron pairs in a thicker high-$Z$
substrate): one then obtains $\eta_\gamma \sim 1.2\,\%$ with a substrate and $\eta_\gamma \simeq 0.9\,\%$ from the uniform target.

%As a conclusion we proved that the substrate present at the back of a nanowire-array target acts as a plasma mirror and can significantly
%rise the efficiency of the high energy photon generation by the synchrotron process. Moreover we demonstrated based on energy-angle
%resolved spectra of photons why the performances of the source are better than in an optimized uniform target.
%and should increase
%the performance of the synchrotron source. Second we report the energy-angle resolved spectra obtained with and without substrate in
%the optimum geometry and compare them to the best uniform target. Our conclusions underline why nanowire-array targets are slightly
%superior in terms of global radiative conversion efficiency as well as the one for a forward cone angle $\pm 30 \ensuremath{^\circ}$. }
%Cette conclusion me semble anticiper un peu trop la conclusion generale qui suit. Evitons donc l'impression de resassement.

%% file: section4_LG.tex
%-----------------------------------------------------------------------
% section 4
%-----------------------------------------------------------------------
\section{Effects of a finite focal spot and an oblique incidence angle}
\label{sec:4} 

All the results of the previous sections correspond to a planar laser wave normally incident on a nanowire array. One may wonder whether they still
hold in the more realistic case of a focused, possibly obliquely incident, laser beam. The variations of the synchrotron yield with the laser incidence
angle have been recently investigated in Ref.~\cite{QESerebryakov2016}, but this study considered planar targets irradiated at a very high laser intensity
($1.3\times 10^{23}\,\mathrm{Wcm}^{-2}$). The strongest emission was found for an incidence angle $\theta_0 \simeq 30\,\%$ and an electron density
$n_{av}\simeq 100n_c$. Our goal here is not to extend this comprehensive study to the case of nanowire arrays but, rather, to examine briefly how the
use of an obliquely incident, focused laser pulse may alter the properties of the emission compared to the optimal planar-wave configuration. To this end,
we have run additional simulations in which the $10^{22}\,\mathrm{Wcm}^{-2}$, $30\,\rm fs$ laser pulse has an 8th-order hyper-Gaussian transverse profile
of FWHM $w=10\,\mu\mathrm{m}$, and impinges onto the target at an angle $\theta_0=0^\circ$ or $30^\circ$. The choice of a hyper-Gaussian transverse
profile aims at minimizing intensity gradient effects, thus easing comparison with the planar-wave results. The laser electric field is in the $xy$ plane
($p$ polarization). The target consists of the highest-yield nanowire setup ($d=36\,\mathrm{nm}$, $D=1\,\mu\mathrm{m}$, $L=10\,\mu\mathrm{m}$ with a Cu
substrate) as previously identified.

\begin{figure}
\centering
\includegraphics[scale=0.8]{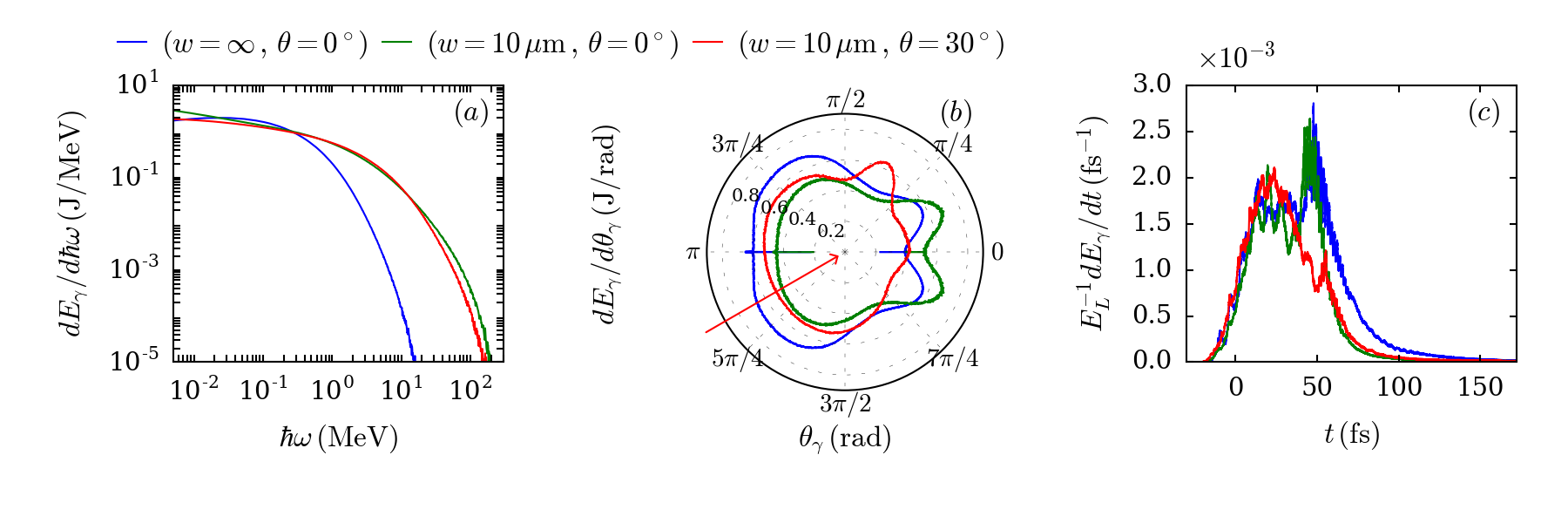}
\caption{Variations of the synchrotron emission with the laser spot size ($w$) and incidence angle ($\theta_0$): (a) energy spectra, (b) angle-resolved
radiated energy and (c) time-resolved radiated power (normalized to the total laser energy $E_L$). Each color represents
a different case as indicated in the legend of panel (a). $w=\infty$ corresponds to the planar wave case. Angles in (b) are defined as
$\theta_\gamma = \arctan (k_{\gamma,y}/k_{\gamma,x}) \in (0, 2\pi)$. The red arrow indicates the $\theta_0=30^\circ$ incidence angle.
All plotted quantities are integrated over the simulation domain.}
\label{fig:figure18a_18c}
\end{figure}

The changes induced by the laser's finite focal spot size and oblique incidence angle on the synchrotron radiation are displayed in
Figs.~\ref{fig:figure18a_18c}(a-c). Since the problem is no longer symmetric relative to the $x$-axis, the photon emission angles are now
defined as $\theta_\gamma = \arctan\left(k_{\gamma,y}/k_{\gamma,x} \right) \in (0,2\pi)$. A striking result [Fig.~\ref{fig:figure18a_18c}(a)]
is that a $10\,\mu\mathrm{m}$ laser focal spot leads to a 10-fold increase in the cutoff photon energy, which attains $\hbar \omega_{\rm max} \simeq 150-180\,\mathrm{MeV}$
(weakly dependent on $\theta_0$) compared to $\hbar \omega_{\rm max} \simeq 16\,\mathrm{MeV}$ for a plane wave. The mean photon energies are also
increased, albeit to a lower extent, from $\langle \hbar \omega \rangle \simeq 2.5\,\mathrm{MeV}$ (above $1\,\mathrm{MeV}$) for a plane wave
to $\langle \hbar \omega \rangle \simeq 3.5\,\mathrm{MeV}$ in the focused case.

\begin{figure}
\centering
\includegraphics[scale=0.8]{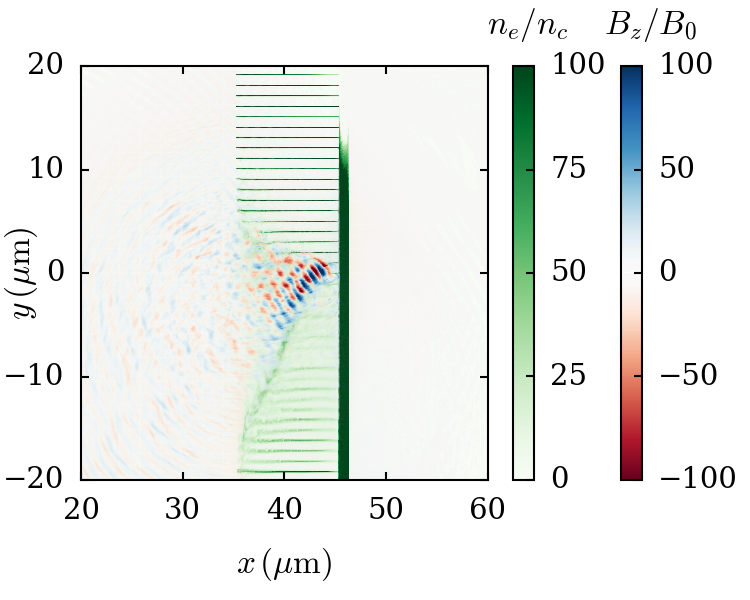}
\caption{Propagation of the obliquely incident, focused laser pulse with $w=10\,\mu\mathrm{m}$ and $\theta_0=30^\circ$: magnetic field ($B_z/B_0$, red-blue colormap)
and electron density ($n_e/n_c$, green colomarp) at $t=83\,\mathrm{fs}$ after the on-target laser peak.}
\label{fig:figure19}
\end{figure}

\begin{figure}
\centering
\includegraphics[scale=0.8]{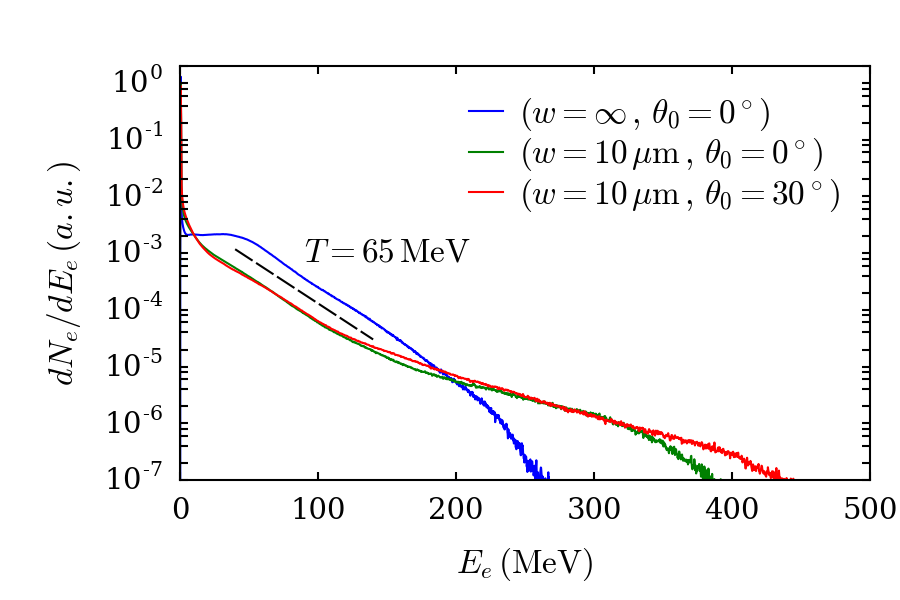}
\caption{Variations of the electron energy spectra $dN_e/dE_e$ with the laser spot size ($w$) and incidence angle ($\theta_0$). The best-fitting
temperature $T=65\,\mathrm{MeV}$ is computed in the $40\le E_e \le 140\,\mathrm{MeV}$ range.}
\label{fig:figure20}
\end{figure}

These enhanced photon energies stem for the relativistic self-focusing undergone by the finite-spot laser pulse in the homogenized plasma \cite{PRLBin2015}.
This phenomenon is illustrated in Fig.~\ref{fig:figure19}, which displays the maps of the magnetic field and electron density in the $\theta_0=30^\circ$ case
at $t=83\,\mathrm{fs}$ after the on-target laser peak. We see that the laser beam has self-focused to a $\sim 2\,\mu\mathrm{m}$ spot where it reaches a maximum
field strength of $B_z/B_0\simeq 100$, consistent with the $\sim 75\,\%$ absorption it has then experienced. The laser self-focusing significantly affects
the electron energy spectra, as shown in Fig.~\ref{fig:figure20} at $t=83\,\mathrm{fs}$. While the electron energy spectra produced by the focused beams show
similar temperatures ($T \simeq 65\,\mathrm{MeV}$, such that $dN_e/dE_e \propto \exp(-E_e/T)$) to the planar wave case up to $E_e \simeq 140\,\mathrm{MeV}$,
they present additional hotter, high-energy tails, extending up to $E_e\simeq 400\,\mathrm{MeV}$. 
 
As a result, the angle-resolved photon spectra obtained with the focused beams show notable differences with the planar-wave case [Fig~\ref{fig:figure18a_18c}(b)].
At $\theta_0=0^\circ$, the backward emission is reduced while the forward radiation is enhanced and emitted into smaller-angle emission lobes
($\theta_\gamma \simeq \pm 20^\circ$). As for the planar wave, the time-resolved radiated power presents two successive maxima corresponding to the laser interaction
with the homogenized nanowires and the substrate [Fig.~\ref{fig:figure18a_18c}(c)]. The overall conversion efficiency is found to be slightly lower than
that observed using a planar wave ($\eta_\gamma \sim 10.2\,\%$ \emph{vs.} $\sim 13\,\%$), with a larger fraction emitted in the forward ($\theta_\gamma <90^\circ$)
direction ($\sim 49\,\%$ \emph{vs.} $\sim 42\,\%$). At $\theta_0=30^\circ$, the backward emission is also lowered (though less than at $\theta_0=0^\circ$), yet the
main difference concerns the forward emission, peaked at angles $\theta_\gamma \simeq 0^\circ$ and $\theta_\gamma \simeq 67^\circ$.
Another difference is the much reduced second maximum in the time-resolved radiated power. This follows from the longer penetration length, and hence
increased absorption of the obliquely propagating laser pulse across the nanowires, which therefore interacts at a lower intensity with the substrate.
This weakened secondary radiation, however, is compensated for by a strengthened radiation throughout the nanowires, thus leading to a total radiation efficiency
($\sim 10.3\,\%$) equal to that obtained at normal incidence.

Overall, those results show that the salient radiation properties evidenced in the planar-wave case are significantly, but not strongly, affected by using a
few nanowires wide focal spot and a moderately oblique incidence angle.

%% file: conclusion_LG.tex
\section*{Conclusions}

Synchrotron radiation from nanowire-array targets irradiated by ultraintense ($10^{21}\le I \le 10^{23}\,\mathrm{Wcm}^{-2}$), ultrashort
($30\,\mathrm{fs}$) laser pulses has been numerically investigated using 2D PIC simulations. We have shown that distinct radiation
mechanisms can arise in the course of the interaction, and at various locations in the target depending on its geometry. A main finding is that,
under the strong-field conditions studied, the wires rapidly expand during the laser pulse, hence forming a fairly (but not fully) homogenized, 
relativistically hot plasma. Consequently, the major emission mechanisms (SDE, RESE, TOEE) previously evidenced in uniform-density
plasmas \cite{PRLRidgers2012, PRLBrady2012, PoPChang2017} also take place in nanowire arrays. Moreover, we have found that
the electrostatic reflection of the energized electrons at the target backside and the slowly-decaying magnetostatic fields induced around
the wires can provide additional radiation channels. In the case of carbon wires driven at $I=10^{22}\,\mathrm{Wcm}^{-2}$, optimum
radiation efficiency, $\eta_\gamma \sim \,\%$ (resp. $\sim 6\,\%$) for $\hbar \omega >10\,\mathrm{keV}$ (resp. $>1\,\mathrm{MeV}$), is
achieved for an interspacing $D\sim 1\,\mu\mathrm{m}$ and wire widths $d\sim 36-50\,\mathrm{nm}$. In the resulting relativistically
transparent ($n_e \sim 20n_c$) plasma, synchrotron emission proceeds through RESE, and is mainly radiated in a large-aperture backward
cone. When increasing the wire width and/or decreasing the interwire spacing so that the homogenized plasma becomes overdense, the
radiation is increasingly forward-directed due to the then-prevailing SDE. Conversely, decreasing the wire width or increasing the interspacing
tends to favor backward radiation processes, \emph{i.e.}, RESE and the backside emission due to refluxing electrons.
Although we have not conducted an extensive parametric scan at laser intensities other than $10^{22}\,\mathrm{Wcm}^{-2}$, we have
found that the radiation efficiency typically increases from $\sim 0.1\,\%$ at $I=10^{21}\,\mathrm{Wcm}^{-2}$ to $\sim 45\,\%$ at
$I=10^{23}\,\mathrm{Wcm}^{-2}$.

While a $\sim 10\,\%$ peak radiation efficiency into $>10\,\mathrm{keV}$ photons is also reported in uniform carbon plasmas of `equivalent' density
($n_{av}=Zn_i d/D$), nanowire arrays are observed to achieve significant ($>3\,\%$) radiation yields up to half solid ($240n_c$) average densities.
Nanowire arrays therefore prove useful not only as practical means of producing, after fast homogenization, plasma targets of controlled sub-solid
density and composition (by varying the array parameters) \cite{PRLBin2015}, but also \emph{per se} as robust and efficient high-energy photon sources.
Besides, we have shown that the radiation yield can be further boosted (up to $\eta_\gamma \sim 13\,\%$) by adding a plasma mirror (a micrometric solid
foil) at the backside of the array. The influence of a finite laser focal spot ($w=10\,\mu\mathrm{m}$) has also been briefly addressed: the increased
laser intensity that results from relativistic self-focusing enhances the production of high-energy electrons, which in turn leads to a slightly 
more forward-directed and significantly more energetic photon source, with a $\times 1.4$ (resp. $\times 10$) increase in the mean (resp. cutoff)
photon energy. Operating with a $30^\circ$ incidence angle gives very similar results. 

To conclude, we remark that our study has assumed a negligible laser prepulse, and therefore that the nanostructure is intact at the arrival of
the intense laser pulse. Were the nanostructure to be prematurely destroyed, it would remain worthwhile to adjust its parameters so as to produce
a relativistically underdense plasma, and to employ a plasma mirror to enhance the total radiation yield. On the theory side, a limitation of
our work, due to computational constraints, is its reduced (2D) geometry. A recent related study \cite{PoPLecz2017} points out that the resonant
processes responsible for electron energization in nanowire arrays may notably differ between 2D and 3D simulations. This work, however, considers
immobile ions and a much weaker intensity ($\sim 6\times 10^{19}\,\mathrm{Wcm}^{-2}$) than ours, so that the array structure is maintained over a
longer time. In our case, by contrast, most of the laser interaction takes place as in a fairly uniformized plasma, which should somewhat mitigate
the 2D/3D discrepancy highlighted in Ref.~\cite{PoPLecz2017}. We therefore expect that our major findings, regarding the nature, efficiency and
interplay of the relevant emission processes, remain mostly valid when moving to 3D. Finally, we neglected the emission of Bremsstrahlung by the
hot electrons, as well as all secondary processes possibly induced by the high-energy photons during their transport, such as electron-positron pair
generation or photonuclear reactions. Some of these limitations will be addressed in future works. 
%While our main findings, regarding the nature, efficiency and interplay
%of the relevant emission processes, should not strongly depend on such simplifications, we expect that moving to 3D~\cite{SABargsten2017}
%and/or focused laser waves could reveal some quantitative differences. 

\ack

One of the authors (E. H) was supported by the French National Research Agency project TULIMA (ANR-17-CE30-0033-01)
and the US Air Force project AFOSR No. FA9550-17-1-0382. The authors acknowledge support by GENCI, France for awarding
up access to HPC resources at TGCC/CCRT (Grant No. 2016-057594 and 2017-057594).

\section*{References}

\bibliographystyle{iopart-num}
\bibliography{myBiblio}